\documentclass{article}

\usepackage[preprint]{neurips_2026}

\usepackage[utf8]{inputenc} %
\usepackage[T1]{fontenc}    %
\usepackage{hyperref}       %
\usepackage{xurl}            %
\usepackage{booktabs}       %
\usepackage{amsfonts}       %
\usepackage{amsmath}        %
\usepackage{nicefrac}       %
\usepackage{microtype}      %
\usepackage[table]{xcolor}  %
\usepackage{graphicx}       %
\usepackage{array}          %
\usepackage{longtable}      %
\usepackage{tabularx}       %
\usepackage{fvextra}
\DefineVerbatimEnvironment{verbatim}{Verbatim}{breaklines=true}
\usepackage{ragged2e}       %
\usepackage{tcolorbox}      %
\newtcolorbox{examplebox}{colback=black!4, colframe=black!70, boxrule=0.6pt, arc=2mm,
  left=6pt, right=6pt, top=4pt, bottom=4pt, fontupper=\small}

\newcommand{\modelname}{Thalamus-Reasoning}

\graphicspath{{figures/}}

\title{\modelname{}: A Multimodal Large Language Model for Multi-Sample Omics Reasoning}

\author{%
  {\bfseries Maciej Sypetkowski,
  Łukasz Smoliński,
  Remigiusz Kinas,
  Joanna Krawczyk} \\
  {\bfseries Bartłomiej Hofman,
  Aleksandra Możwiłło,
  Michał Pikusa,
  Tomasz Jetka,
  Rafał Powalski}\thanks{Corresponding author: \texttt{rafal.powalski@ingenix.ai}} \\
  Ingenix.ai
}

\begin{document}

\raggedbottom

\maketitle

\begin{abstract}
Large language models can reason over biological knowledge, but they are not naturally equipped to work with quantitative omics measurements. Existing approaches either convert expression profiles into long textual representations, often losing quantitative information, or connect omics encoders to language models for narrow tasks. 
We introduce \modelname{}, a 9B multimodal language model that represents each bulk or single-cell transcriptome as a compact continuous embedding that can be interleaved with natural-language instructions and with other samples in the same context. 
A key contribution is a training corpus that connects experimental measurements with biological knowledge and task supervision. We combine templates linking transcriptomic profiles to annotations and experimental outcomes, reasoning examples connecting evidence across sources, and LLM-generated supervision with task-specific quality checks. The instruction-tuning corpus spans 157 biological and general-purpose task groups. Training includes continued pretraining on 20 billion tokens and supervised fine-tuning on 5.21 million conversations. We further introduce GEO-OmicsQA, a publication-disjoint benchmark of 3,000 questions requiring models to interpret and compare expression profiles from real GEO studies. 
\modelname{} achieves the best results among the evaluated models on three of four held-out annotation tasks and on all six GEO-OmicsQA variants, and the best average performance among compared methods on PerturbQA perturbation-response prediction. Ablations show a clear advantage for continuous omics representations over a text-only gene-list representation. Together, these results show that quantitative transcriptomic measurements and biological language can be integrated within a general conversational model, enabling both predictive tasks and language-guided interpretation of experimental samples.
\end{abstract}

\begin{figure}[t]
  \centering
  \includegraphics[width=\linewidth]{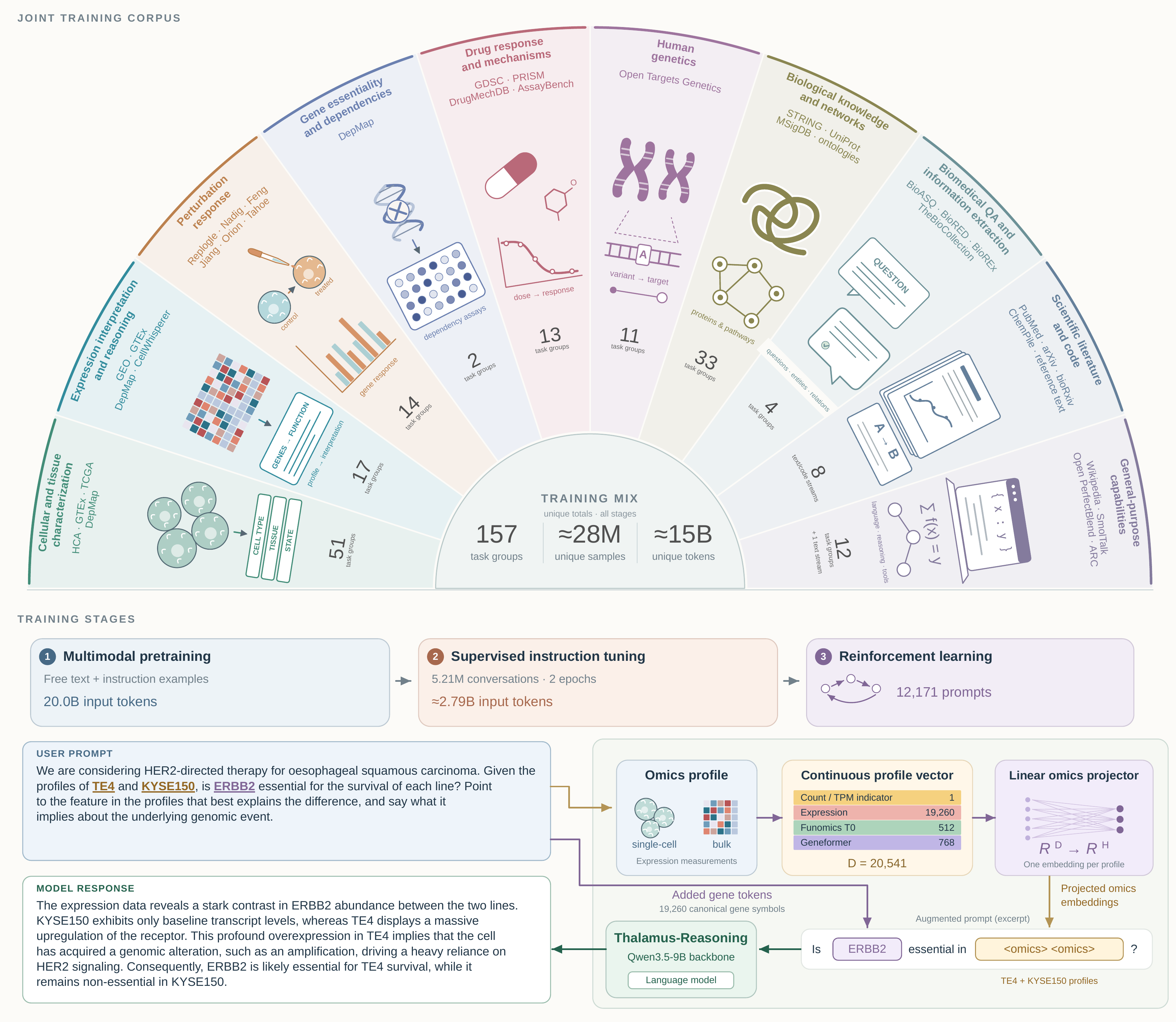}
  \caption{\textbf{\modelname{} overview.}
  \textit{Top:} the joint training corpus spans ten capability groups. Sector counts regroup the 157 instruction-tuning task groups in Table~\ref{tab:task_examples}; scientific text/code streams and the Wikipedia text stream are counted separately. The centre gives their total and the unique samples and tokens seen across all training stages: pretraining counts only records actually sampled (large text sources such as arXiv were only partly used), and examples repeated across epochs or stages are counted once. \textit{Middle:} the three training stages comprise multimodal pretraining, supervised instruction tuning, and reinforcement learning. \textit{Bottom:} single-cell or bulk expression profiles form continuous vectors projected into the language-model input at \texttt{<omics>} positions, alongside text and added gene tokens. The example conversation at lower left connects to the architecture at lower right; highlighted cell names are augmented with profile embeddings. Sector areas are schematic.}
  \label{fig:overview}
\end{figure}

\section{Introduction}
\label{sec:intro}

The advent of high-throughput sequencing technologies has revolutionized our understanding of disease progression, tissue development, and cellular heterogeneity by providing quantitative measurements of biological molecules at both bulk and single-cell resolutions \citep{regev2017science, eberwine2014single}. Transcriptomic profiles are among the most informative of these measurements: a single RNA-seq assay quantifies the expression of thousands of genes, reflecting cell identity and state as well as responses to disease, genetic perturbation, and drug treatment. Public atlases and repositories now provide such profiles across tissues, cell types, diseases, and experimental conditions \citep{regev2017science, gtex2020atlas, weinstein2013tcga, clough2024geo}, making expression a natural interface between experimental measurements and biological knowledge. Concurrently, large language models (LLMs) have demonstrated unprecedented capabilities in reasoning, knowledge retrieval, and instruction following \citep{brown2020language, openai2023gpt4}. However, a significant gap remains between the quantitative, high-dimensional nature of omics data and the text-centric input space of modern LLMs.

Existing biological foundation models tackle parts of this gap: specialized single-cell models learn representations for downstream prediction \citep{cui2024scgpt, theodoris2023geneformer, hao2024scfoundation, rosen2023uce}, while text-based biomedical LLMs serialize expression as gene-name sequences \citep{levine2024cell2sentence, rizvi2025scaling, choi2026pgl}, requiring long inputs and, for rank-only representations, discarding expression magnitudes. Multimodal models connect quantitative transcriptomic representations to language \citep{schaefer2025cellwhisperer, fang2025instructcell}.

Many biological questions are comparative: control versus perturbation, patient versus model, sensitive versus resistant, or one cellular state versus another. Long textual representations consume context as more profiles are added, and interfaces designed primarily around individual profiles may require adaptation for joint comparisons. These constraints can make multi-sample analysis cumbersome. We ask whether quantitative transcriptomes can instead be represented compactly inside a language model, allowing several experimental samples to be analyzed jointly with biological context and natural-language instructions. Learning to make these comparisons also requires suitable supervision connecting experimental measurements to biological knowledge, task instructions, and examples of interpreting evidence across samples.

In this work, we introduce \textbf{\modelname{}}, a multimodal language model trained for diverse omics tasks (Figure~\ref{fig:overview}). We adopt the standard modality-projection approach used in vision-language models \citep{liu2024visual, bai2023qwenvl} and find it effective for transcriptomic inputs. A linear layer maps normalized expression values together with pretrained omics embeddings into the input embedding space of an LLM. Each profile occupies a compact representation that can be interleaved with natural-language instructions and other profiles, enabling direct comparisons across experimental samples. We also augment the vocabulary with explicit gene tokens to represent biological nomenclature efficiently.

To help the model interpret omics data in its biological context, we devote substantial effort to constructing the training mixture: templates linking experimental measurements to curated biological knowledge, reasoning examples connecting evidence across sources, and LLM-generated supervision. It draws on expression atlases, perturbation screens, human genetics, and scientific literature. We organise the biological and general-purpose supervision used for instruction tuning into 157 task groups by biological objective and input/output format, bundling related cohorts, context variants, and prompt variants (Table~\ref{tab:task_examples}). Training proceeds through 20 billion input tokens of continued pretraining, supervised fine-tuning on 5.21 million conversations (3.49 million with omics inputs), and reinforcement learning.

For context, C2S-Scale reports 12 pretraining task formulations \citep{rizvi2025scaling}, InstructCell three biological tasks \citep{fang2025instructcell}, and CellWhisperer a generated pool of 106{,}610 chats \citep{schaefer2025cellwhisperer}. Our 157 task groups and 5.21 million SFT conversation records include both biological and general-purpose supervision. Although counting conventions differ, our corpus substantially expands the volume and diversity of supervision used for omics--language training. Sections~\ref{sec:training_procedure} and~\ref{sec:datasets} describe the training stages and data construction.

Our main contributions are:
\begin{itemize}
    \item \textbf{\modelname{}}, a conversational model, fusing continuous transcriptomics with language, supporting several samples in context, evaluated on annotation, perturbation prediction, and biological question answering.
    \item \textbf{GEO-OmicsQA}, a benchmark for language-guided interpretation of one or more transcriptomic profiles.
    \item \textbf{Large-scale construction of omics--language training data}, transforming heterogeneous experimental measurements, curated knowledge, and scientific literature into supervision for interpreting biological profiles. We combine data harmonization, task design, and evidence-based filtering of generated reasoning to support a training procedure spanning pretraining, instruction tuning, and reinforcement learning.

\end{itemize}

\section{Related Work}
\label{sec:related}

General-purpose Gemini \citep{google2026gemini31pro, google2026gemini38flash}, GPT \citep{openai2026gpt6soluna}, and Qwen \citep{qwen2026qwen359b, qwen2026qwen38} models offer broad knowledge and reasoning, but dedicated training on quantitative omics measurements paired with biological tasks is not established in their published training descriptions.

\paragraph{Omics Foundation Models}
scGPT \citep{cui2024scgpt}, Geneformer \citep{theodoris2023geneformer, chen2026geneformer}, and scFoundation \citep{hao2024scfoundation} learn gene and cell representations through self-supervised transcriptomic pretraining. Funomics extends representation learning across bulk and single-cell expression \citep{funomics}, while GenePT derives embeddings from textual gene descriptions \citep{chen2025genept}. These features support downstream prediction and can serve as inputs to instruction-following generative models.

\paragraph{Text-Based Biomedical LLMs}
Cell2Sentence (C2S) \citep{levine2024cell2sentence} and C2S-Scale \citep{rizvi2025scaling} transform expression matrices into ``cell sentences''---plain-text strings of gene names ordered by descending abundance. Portraying Gene Language (PGL) \citep{choi2026pgl} explores transcriptome generation from metadata using gene-expression token sequences. Cell-o1 \citep{fang2025cello1} similarly represents each cell with ranked top-expressed genes and donor metadata, training a 7B LLM with supervised fine-tuning and reinforcement learning for batch-level annotation. While compatible with standard language-model architectures, rank-only representations omit explicit expression magnitudes, and longer gene lists consume context that could otherwise accommodate additional samples or instructions.
Beyond expression serialization, CellHermes \citep{gao2025cellhermes} converts transcriptomic and protein-interaction data into templated question--answer pairs for joint training. For perturbation prediction, SUMMER \citep{wu2025perturbqa} combines biological knowledge summaries with retrieval of experimental outcomes, and SynthPert \citep{phillips2025synthpert} trains on synthetic mechanistic rationales conditioned on known outcomes and filtered by a critic.

\paragraph{Multimodal Omics-Language Alignment}
CellWhisperer \citep{schaefer2025cellwhisperer} combines contrastive transcriptome--text alignment with a generative chat model that projects transcriptome embeddings into the LLM token space. It uses paired profiles and descriptions for alignment and generated conversations for chat supervision. InstructCell \citep{fang2025instructcell} pairs quantitative cell inputs with language instructions, constructing supervision from biological attributes and LLM-generated instruction--response templates. Related interfaces include scMMGPT's cross-modal alignment and generation objectives \citep{shi2025scmmgpt} and Cell2Text's adaptation of gene-level embeddings for cell-description generation \citep{kharouiche2025cell2text}. AROMA \citep{wang2026aroma} combines retrieved text with graph and protein representations for perturbation reasoning. We build on these interfaces with a training mixture covering diverse tasks and comparative interpretation of multiple samples.

\paragraph{Biological Question Answering Benchmarks}
CellVerse \citep{zhang2025cellverse}, SOAR \citep{liu2024soar}, and scBench \citep{workman2026scbench} cover cell-type annotation, perturbation analysis, scientific QA, and practical single-cell workflows. SC-Arena \citep{zhao2026scarena} extends natural-language evaluation to cell captioning, generation, perturbation prediction, and scientific QA; PerturbQA \citep{wu2025perturbqa} tests experimental perturbation outcomes and gene-set interpretation. GEO-OmicsQA complements these benchmarks with publication-disjoint evaluation of language-guided interpretation across real expression profiles.

\section{Method}
\label{sec:method}

The \modelname{} architecture is designed to bridge the modality gap between continuous omics data and discrete natural language. By treating single-cell and bulk transcriptomes as a continuous modality analogous to visual inputs in vision-language models (VLMs), we enable a pretrained LLM to condition on high-dimensional biological states. In this section, we describe the model architecture and training procedure; the training corpus is described in Section~\ref{sec:datasets}, and the GEO-OmicsQA benchmark in Section~\ref{sec:results_geo_qa}.

\subsection{Model Architecture}
\label{sec:method_arch}

\paragraph{Composite Omics Representations}
\modelname{} represents each expression profile as a continuous vector. We align all expression profiles to a fixed human gene panel of $G = 19{,}260$ protein-coding and mitochondrial genes derived from GENCODE annotations \citep{mudge2025gencode}. For each sample, the default input vector concatenates four feature groups: a one-dimensional input-scale indicator, the normalized expression vector, a 512-dimensional Funomics T0 embedding \citep{funomics}, and a 768-dimensional embedding from Geneformer-V2-104M \citep{chen2026geneformer}, yielding a 20{,}541-dimensional omics vector. This combination gives the language model access to gene-level expression magnitudes alongside representations learned from transcriptomic data. The Funomics and Geneformer embeddings are precomputed with frozen encoders, which are not updated during language-model training.
The input-scale indicator distinguishes count-derived single-cell profiles from transcripts-per-million (TPM)-normalized bulk RNA-seq profiles, aligning the expression values with their source modality.

The feature groups are normalized before training to make their scales comparable. Normalization statistics are computed from the training set. Expression values are first library-size normalized to $10^6$, log-transformed with $\log(1+x)$, centered per gene, and scaled by a single standard deviation computed over the full expression range. The input-scale indicator and learned embedding features are centered and scaled per dimension.

\paragraph{Cross-Modal Alignment via Linear Projection}
We use the text backbone of Qwen3.5-9B \citep{qwen2026qwen35} as the language model. To connect the omics vector to this backbone, \modelname{} adds a trainable affine projection
\[
    \mathcal{P}_{\theta}(\mathbf{v}) = W\mathbf{v} + \mathbf{b}, \qquad
    \mathbf{v} \in \mathbb{R}^{D}, \quad W \in \mathbb{R}^{H \times D},
\]
where $D=20{,}541$ is the dimensionality of the full omics input vector and $H$ is the LLM token-embedding size. The projection uses a bias term, is initialized with Xavier-uniform weights scaled by gain 0.01 and zero bias, and is trained jointly with the language model.

\modelname{} reuses the multimodal placeholder mechanism provided by the Qwen3-VL template \citep{bai2025qwen3vl}, with the omics projection supplying the modality embeddings. We denote each profile placeholder schematically as \texttt{<omics>}. In the model input, each profile occupies one projected embedding $\mathcal{P}_{\theta}(\mathbf{v}_j)$ bracketed by two modality delimiter tokens. This lets omics profiles enter the LLM as continuous token embeddings while preserving the surrounding natural-language prompt. Formally, for text tokens $x_i$ and omics vectors $\mathbf{v}_j$, the input embedding sequence, omitting the delimiters for clarity, becomes
\[
    [E_{\text{txt}}(x_1), \ldots, \mathcal{P}_{\theta}(\mathbf{v}_1), \ldots,
    E_{\text{txt}}(x_k), \mathcal{P}_{\theta}(\mathbf{v}_2), \ldots].
\]
Because each profile appears at a designated position in the input sequence, a prompt can interleave multiple biological samples with natural-language instructions while preserving most of the context window for the task description and generated answer.

\paragraph{Gene-Aware Vocabulary Augmentation}
Standard LLM tokenizers are optimized for common human languages and often fragment biological nomenclature. A gene symbol such as \texttt{TP53} or \texttt{BRCA1} may be split into several subword pieces, making gene-level generation and matching unnecessarily brittle. We therefore augment the base tokenizer with the same canonical human gene symbols used for the 19{,}260-gene expression panel. For a gene symbol $g$ that originally tokenizes to subword IDs $T(g)$, the default initialization sets the new token embedding to
\[
    E_{\text{new}}(g) = \frac{1}{|T(g)|}\sum_{t \in T(g)} E_{\text{old}}(t).
\]
This initializes gene tokens from the base model's subword embeddings and allows canonical gene names to be read and generated as atomic tokens. The added gene embeddings are trainable together with the rest of the language model. Encoding each of the 19{,}260 gene symbols independently with the unmodified Qwen3.5 tokenizer, without leading whitespace or special tokens, requires an average of $3.72$ tokens per gene. Vocabulary augmentation reduces this to one token per gene, shortening prompts and target sequences that mention many genes.

\subsection{Training Procedure}
\label{sec:training_procedure}

We jointly update the language backbone, token embeddings, and omics projection while keeping the omics encoders frozen. The three stages below use the data described in Section~\ref{sec:datasets}; optimization settings and compute requirements are in Appendix~\ref{app:training_details}.

\paragraph{Stage 1: Multimodal Pretraining}
Starting from the augmented pretrained model, we train on 20 billion input tokens, interleaving continued pretraining (CPT) on free text with supervised fine-tuning (SFT) on instruction examples. This stage connects biological knowledge with quantitative omics inputs through alignment and prediction tasks. The causal language-modeling objective supervises text in free-text records and assistant responses in conversational records. Source sampling weights balance exposure across datasets (Appendix~\ref{app:pretraining}).

\paragraph{Stage 2: Supervised Instruction Tuning}
Starting from Stage 1, we concentrate training on 5.21 million biological and general-purpose conversations, including direct answers, reasoning traces, and multi-turn interactions. We minimize negative log-likelihood on selected assistant responses; other messages and excluded assistant turns remain context without contributing to the loss. Appendix~\ref{app:sft_training} details the corpus and supervision.

\paragraph{Stage 3: Reinforcement Learning}
We refine the SFT model with group relative policy optimization (GRPO; \citealp{shao2024deepseekmath}) in VERL~\citep{sheng2024hybridflow}, optimizing sampled responses against deterministic task verifiers. Difficulty filtering with the SFT initializer selects 12{,}171 prompts that it can solve occasionally, excluding unsuccessful and nearly saturated groups. Rewards primarily measure task outcomes, with smaller terms for response structure and evidence agreement where ledgers are available. A KL penalty to the frozen SFT reference limits policy drift. Appendix~\ref{app:rl_training} gives filtering coverage, reward definitions, optimization settings, and exploratory small-model results, which include regressions as well as gains.

\section{Training Data}
\label{sec:datasets}

The training mixture connects single-cell and bulk measurements with biological knowledge and reasoning, complemented by scientific text and general-purpose instructions. We group subsets into ten broad corpus categories by their primary purpose; one source may contribute to several categories. These categories describe the overall training mixture, whereas Table~\ref{tab:task_examples} provides a finer grouping of Stage 2 supervision into 157 task groups. Source preparation and inventories are documented in Appendix~\ref{app:datasets}, with representative tasks in Appendix~\ref{app:task_examples}.
\begin{itemize}
    \item \textbf{Cellular characterization:} Predicting cell type, tissue, lineage, cell-cycle phase, malignancy, clinical attributes, and cancer-model properties from expression profiles.
    \item \textbf{Expression interpretation and reasoning:} Identifying expressed or silent genes, comparing samples, describing profiles, interpreting pathway and interaction-partner activity, and answering profile-grounded questions.
    \item \textbf{Perturbation response:} Predicting differential expression, response direction and affected genes; identifying interventions; comparing doses; and transferring responses across cellular contexts.
    \item \textbf{Gene essentiality and functional screens:} Predicting context-specific essentiality and identifying or ranking gene hits in phenotypic screens.
    \item \textbf{Drug response and mechanisms:} Predicting sensitivity and quantitative response endpoints, identifying targets, and explaining drug mechanisms and cellular responses.
    \item \textbf{Human genetics:} Assessing gene--disease evidence, classifying variant consequences, and prioritizing candidate genes at GWAS loci.
    \item \textbf{Biological knowledge and networks:} Cell-line genomic facts, protein properties, pathway and interaction-graph reasoning, gene-set membership, and cell and disease ontologies.
    \item \textbf{Biomedical QA and information extraction:} Text-only biomedical questions, answers, reasoning traces, and entity-relation extraction.
    \item \textbf{Scientific and biomedical text and code:} Literature, clinical-trial records, drug labels, research-project descriptions, and biomedical code.
    \item \textbf{General-purpose capabilities:} General text, instruction following, mathematics, coding, multi-turn dialogue, system-instruction following, and tool use.
\end{itemize}

We use contextual cell-line detection to link mentions in text with DepMap reference profiles, interleaving biological text and quantitative measurements in the same input sequence (Appendix~\ref{app:cell_line_augmentation}).

Constructing several of these datasets required multistage integration of experimental measurements, curated biological knowledge, and literature evidence, together with quality checks on the generated explanations. The drug mechanism-of-action and marker-grounded reasoning datasets illustrate this process; their construction and filtering procedures are detailed in Appendices~\ref{app:moa_dataset} and~\ref{app:markers_cot}.

\paragraph{Benchmark exclusions.}
We apply benchmark-specific exclusions during training-data construction. GEO-OmicsQA test studies and their associated publications are withheld from training; Tabula Sapiens is reserved for evaluation; and 5{,}683 Human Diseases samples overlapping with the GEO training collection are removed. For PerturbQA, held-out cell-line--perturbation pairs are excluded from genetic training screens and derived task templates. Appendices~\ref{app:datasets}, \ref{app:geo_omicsqa_protocol}, and~\ref{app:perturbation_prediction} describe the scope of these exclusions.

\section{Results}
\label{sec:results}
We evaluate \modelname{} across three complementary biological capabilities: identifying samples, interpreting experimental comparisons, and predicting perturbation responses. First, zero-shot annotation tests whether the model can recover cell type, tissue, and disease identity from expression profiles in held-out datasets. Second, GEO-OmicsQA evaluates language-guided interpretation of real experimental profiles, including comparisons across multiple samples. Third, PerturbQA tests whether the same model can predict the direction of gene-expression changes following a perturbation from an unperturbed control profile and biological context. We then use ablations to examine how input representations, backbone size, and gene tokenization contribute to performance.

\subsection{Zero-Shot Annotation}
\label{sec:results_zero_shot_annotation}
We first test whether \modelname{} can identify the biological properties of samples from their expression profiles in annotation datasets unseen during training. We use Tabula Sapiens, which is excluded entirely from training, and Human Diseases, whose overlapping samples were removed from the GEO training corpus (Table~\ref{tab:datasets}). The tasks cover cell type and tissue of origin in Tabula Sapiens, and disease subtype and tissue in Human Diseases.

We compare with CellWhisperer, C2S-Scale-27B, and Cell-o1 using macro AUROC. For \modelname{}, C2S-Scale, and Cell-o1, candidate labels are supplied in the prompt and scored by length-normalized constrained likelihood (Appendix~\ref{app:constrained_likelihood}); C2S-Scale and Cell-o1 receive the 1{,}000 highest-expressed genes as text, while \modelname{} receives omics embeddings. We evaluate CellWhisperer's Mistral-7B chat model using its transcriptome embeddings and constrained likelihood scoring with task-specific answer prefixes. We also retain its published Geneformer--BioBERT embedding-model results, denoted CellWhisperer (CLIP).

CellWhisperer (LM) achieves the highest macro AUROC on Tabula Sapiens cell type top-20, while \modelname{} leads on Tabula Sapiens tissue and both Human Diseases tasks (Table~\ref{tab:tabula_geo_disease}). The transcriptome-embedding models outperform both ranked-gene baselines. C2S-Scale's poor Human Diseases disease-subtype result may partly reflect task mismatch: its reported pretraining objectives include cell-type and tissue annotation, but not disease-subtype classification \citep{rizvi2025scaling}.

\begin{table}[!htbp]
  \small
  \caption{\textbf{Tabula Sapiens and Human Diseases classification zero-shot benchmarks.}
  Entries are macro AUROC. CellWhisperer (CLIP) reports published embedding-model results; CellWhisperer (LM) is our evaluation of its released chat model. All language-model rows use constrained likelihood scoring over the candidate labels.}
  \label{tab:tabula_geo_disease}

  \centering
  \resizebox{\linewidth}{!}{%
  \begin{tabular}{lrrrr}
    \toprule
    Model & \multicolumn{2}{c}{Tabula Sapiens} & \multicolumn{2}{c}{Human Diseases} \\
    \cmidrule(lr){2-3}\cmidrule(lr){4-5}
    & Cell type top-20 & Tissue & Disease subtype & Tissue \\
    \midrule
    \modelname{} & 0.926 & \textbf{0.767} & \textbf{0.906} & \textbf{0.984} \\
    CellWhisperer (LM) \citep{schaefer2025cellwhisperer} & \textbf{0.948} & 0.760 & 0.839 & 0.932 \\
    CellWhisperer (CLIP) \citep{schaefer2025cellwhisperer} & 0.942 & 0.747 & 0.819 & 0.872 \\
    C2S-Scale-27B \citep{rizvi2025scaling} & 0.684 & 0.592 & 0.543 & 0.631 \\
    Cell-o1 \citep{fang2025cello1} & 0.851 & 0.720 & 0.706 & 0.820 \\
    \bottomrule
  \end{tabular}}
\end{table}

\subsection{GEO-OmicsQA}
\label{sec:results_geo_qa}
We next move from assigning sample labels to interpreting experimental profiles through natural-language questions, including comparisons across samples. We introduce \textbf{GEO-OmicsQA} to evaluate this capability using one or more expression profiles and natural-language context. GEO~\citep{clough2024geo} is a large, community-deposited archive of real-world studies from independent laboratories, spanning tissues, primary cells, cell lines, diseases, and genetic or chemical perturbations, each with its own experimental design and unstructured, study-specific metadata (Appendix~\ref{app:datasets}). GEO-OmicsQA inherits this heterogeneity: its 3{,}000 questions come from 235 studies and 183 publications, drawn from a held-out pool of 879 studies and 793 publications, and each is generated from its study's publication, so it refers to the samples, conditions, and findings of that specific study. Questions range from yes/no checks of reported findings to open-ended explanations of expression differences between conditions, and they are clustered by study, with up to 57 questions from a single study. The benchmark comprises 1{,}000 balanced binary questions, 1{,}000 simple free-text questions, and 1{,}000 contextual free-text questions with auxiliary literature context. Each family has 500 standard and 500 minimal prompts. Test studies are publication-disjoint from the training data; a subset was reviewed by biological experts for question--answer validity, sample grounding, and biological coherence. Construction details and evidence annotations are provided in Appendix~\ref{app:geo_omicsqa_protocol}.

Minimal prompts reduce textual scaffolding while retaining the same task and answer space. The standard and minimal subsets are not fully paired, so score differences may also reflect question composition.

We run nine language baselines with ranked lists of up to 1{,}000 genes per sample, adapting the count to each prompt's token budget; \modelname{} receives matched omics embeddings. We omit CellWhisperer from this benchmark because adapting its released single-profile chat interface to these multi-sample prompts is outside the scope of this comparison. Binary answers use balanced yes/no accuracy and free-text answers use mean DeepEval judge score (Appendix~\ref{app:deepeval_judge}). Input budgets, decoding settings, and C2S-Scale's yes/no logit scoring are detailed in Appendix~\ref{app:geo_omicsqa_protocol}.

\modelname{} achieves the highest score across all six prompt variants (Table~\ref{tab:geo_qa}), with aggregate binary accuracy of 0.767 and a free-text judge score of 0.651, compared with 0.662 and 0.613 for Gemini~3.8 Flash. Although ranked gene lists are not a native quantitative input to general-purpose LLMs, several models make effective use of this representation: Flash leads the baselines on every variant, followed by Pro and the GPT models. The very low C2S-Scale free-text score reflects the released checkpoint's failure to answer questions under this protocol (Appendix~\ref{app:c2s_geoqa_behavior}).

\begin{table}[!htbp]
  \scriptsize
  \caption{\textbf{GEO-OmicsQA benchmark.}
  Aggregate columns average over their corresponding variants. Binary QA reports balanced \emph{yes}/\emph{no} accuracy, while free-text reports Gemini-judge score (full DeepEval judge details are in Appendix~\ref{app:deepeval_judge}).}
  \label{tab:geo_qa}

  \centering
  \setlength{\tabcolsep}{3pt}
  \resizebox{\linewidth}{!}{%
  \begin{tabular}{lrr@{\hspace{0.6em}}rrrrrr}
    \toprule
    & \multicolumn{2}{c}{Aggregate} & \multicolumn{6}{c}{By prompt variant} \\
    \cmidrule(lr){2-3}\cmidrule(lr){4-9}
    Model & Binary QA & Free text & QA std. & QA min. & Simple min. & Simple std. & Context min. & Context std. \\
    \midrule
    \multicolumn{9}{l}{\textit{Ours}} \\
    \modelname{} & \textbf{0.767} & \textbf{0.651} & \textbf{0.752} & \textbf{0.782} & \textbf{0.560} & \textbf{0.672} & \textbf{0.638} & \textbf{0.734} \\
    \midrule
    \multicolumn{9}{l}{\textit{General LLMs}} \\
    Gemini 3.8 Flash \citep{google2026gemini38flash} & 0.662 & 0.613 & 0.712 & 0.612 & 0.512 & 0.636 & 0.607 & 0.696 \\
    Gemini 3.1 Pro Preview \citep{google2026gemini31pro} & 0.640 & 0.536 & 0.690 & 0.590 & 0.442 & 0.544 & 0.546 & 0.612 \\
    GPT-6 Sol \citep{openai2026gpt6soluna} & 0.606 & 0.510 & 0.632 & 0.580 & 0.438 & 0.525 & 0.517 & 0.561 \\
    GPT-6 Luna \citep{openai2026gpt6soluna} & 0.575 & 0.479 & 0.588 & 0.562 & 0.411 & 0.474 & 0.495 & 0.537 \\
    Qwen3.5-9B \citep{qwen2026qwen359b} & 0.534 & 0.360 & 0.542 & 0.526 & 0.319 & 0.355 & 0.369 & 0.395 \\
    Qwen3.8-27B \citep{qwen2026qwen38} & 0.546 & 0.403 & 0.560 & 0.532 & 0.351 & 0.397 & 0.408 & 0.457 \\
    \midrule
    \multicolumn{9}{l}{\textit{Biology-specialized models}} \\
    TxGemma-27B \citep{wang2025txgemma} & 0.518 & 0.289 & 0.532 & 0.504 & 0.222 & 0.282 & 0.307 & 0.344 \\
    Cell-o1 \citep{fang2025cello1} & 0.522 & 0.415 & 0.524 & 0.520 & 0.330 & 0.431 & 0.402 & 0.496 \\
    C2S-Scale-27B \citep{rizvi2025scaling} & 0.521 & 0.028 & 0.514 & 0.528 & 0.023 & 0.024 & 0.033 & 0.031 \\
    \bottomrule
  \end{tabular}}
\end{table}

\subsection{Perturbation Prediction}
\label{sec:results_perturbation}

Finally, we evaluate prediction of a perturbation response from an unperturbed profile and biological context. On PerturbQA~\citep{wu2025perturbqa}, we predict whether a readout gene increases or decreases after CRISPRi of a target gene, conditional on the readout being differentially expressed. We use the official direction-of-change TEST split: 16{,}208 questions across four cell lines. \modelname{} receives the unperturbed control profile and public gene-pair notes, and generates an Increase/Decrease answer.

We rescore released SUMMER, retrieval, GenePT, and GEARS~\citep{roohani2024gears} predictions and reproduce AROMA on the same test identities. SynthPert is included only as a published reference because its trained checkpoint is unavailable. Table~\ref{tab:perturbqa_direction} defines the aggregation and metrics; Appendix~\ref{app:perturbation_prediction} details scoring and reproduction checks.

\begin{table}[!ht]
  \centering
  \caption{\textbf{PerturbQA direction-of-change prediction}~\citep{wu2025perturbqa}. AUROC is averaged over readout genes within each cell and then equally over cells; MF1 is cell-averaged direction macro-F1. All locally scored rows use the same 16{,}208 TEST questions; \modelname{} and AROMA average three sampled answer scores per question. Bold denotes the best of these scores. SynthPert (paper) is a published reference with unverified test coverage; its mean averages rounded paper values. For SUMMER, retrieval, GenePT, and GEARS, macro-F1 uses released scores thresholded at 0.5; the original decision rules are unverified. --- denotes unavailable F1.}
  \label{tab:perturbqa_direction}
  \small
  \setlength{\tabcolsep}{4pt}
  \begin{tabular}{lrrrrrr}
    \toprule
    & \multicolumn{5}{c}{AUROC} & \\
    \cmidrule(lr){2-6}
    Model & K562 & RPE1 & HepG2 & Jurkat & Mean & MF1 \\
    \midrule
    \modelname{} & \textbf{.7939} & \textbf{.7173} & \textbf{.7679} & \textbf{.7750} & \textbf{.7635} & \textbf{.8761} \\
    SUMMER~\citep{wu2025perturbqa} & .6379 & .6426 & .6634 & .6748 & .6547 & .7942 \\
    Retrieval without LLM~\citep{wu2025perturbqa} & .6913 & .6331 & .6380 & .6098 & .6431 & .7875 \\
    GenePT protein~\citep{chen2025genept} & .5583 & .5857 & .5558 & .5986 & .5746 & .7447 \\
    GenePT gene~\citep{chen2025genept} & .4891 & .5891 & .6112 & .5828 & .5681 & .7128 \\
    GEARS~\citep{roohani2024gears} & .6660 & .6064 & .5195 & .5181 & .5775 & .4648 \\
    AROMA~\citep{wang2026aroma} & .5645 & .5683 & .5987 & .5838 & .5788 & .6934 \\
    \midrule
    SynthPert (paper)~\citep{phillips2025synthpert} & .65 & .73 & .72 & .65 & .6875 & --- \\
    \bottomrule
  \end{tabular}
\end{table}

\modelname{} attains 0.7635 mean AUROC, exceeding released SUMMER by 0.1088, with higher AUROC in every cell line (Table~\ref{tab:perturbqa_direction}). Its direction macro-F1 is 0.8761, compared with SUMMER's diagnostic 0.7942. Detailed results are in Appendix~\ref{app:perturbation_prediction}. These descriptive comparisons involve systems with different inputs, training data, and inference budgets; they do not isolate omics conditioning or establish statistical significance. We make no unseen-cell-line or globally unseen-target generalization claim.

\subsection{Ablations}
\label{sec:results_ablations}
We examine the effects of backbone size, omics inputs, and training stages
(Table~\ref{tab:ablation_compact}). Appendix~\ref{app:ablations} reports per-task
scores, variability across runs, and additional input and gene-token ablations.

\begin{table}[t]
\centering\small
\caption{\textbf{Selected ablation results.} Overall is the unweighted mean of seven benchmark scores on a 0--100 scale (higher is better). The reference configuration uses Qwen3.5-0.8B, the full omics input, gene tokens, and Pretraining $\rightarrow$ SFT. Scores average three independent training runs; both 9B configurations report a single run. $\dagger$ denotes the final \modelname{} model. Bold denotes the best score in each block.}
\label{tab:ablation_compact}
\resizebox{\linewidth}{!}{%
\begin{tabular}{@{}lr@{\hskip 1.2em}lr@{\hskip 1.2em}lr@{\hskip 1.2em}lr@{}}
\toprule
\multicolumn{2}{c}{\textit{Backbone size}} & \multicolumn{2}{c}{\textit{Omics input}} & \multicolumn{2}{c}{\textit{Training stages}} & \multicolumn{2}{c}{\textit{Reinforcement learning}} \\
\cmidrule(r){1-2}\cmidrule(lr){3-4}\cmidrule(lr){5-6}\cmidrule(l){7-8}
0.8B & 77.6 & Full input & \textbf{77.6} & Pretraining $\rightarrow$ SFT & \textbf{77.6} & 9B & 81.9 \\
2B & 79.9 & No expression & 76.2 & Pretraining only & 75.3 & 9B + RL$^{\dagger}$ & \textbf{82.4} \\
9B & \textbf{81.9} & Cell2Sentence-like & 71.5 & SFT only & 73.4 & & \\
\bottomrule
\end{tabular}}
\end{table}

The ablations support three conclusions about building a shared model for omics
and language. First, compact continuous profiles outperform the tested
ranked-gene representation, supporting their use as an interface to experimental
measurements. Second, combining pretraining with instruction tuning yields
stronger aggregate performance than either stage alone, supporting the staged
training recipe. Third, larger backbones improve biological question answering
and perturbation prediction more consistently than sample annotation, suggesting
that additional capacity benefits these capabilities differently. Reinforcement
learning provides a smaller further improvement. The resulting picture is one of
complementary contributions from the input representation, training procedure,
and model capacity.

\section{Conclusion}
\label{sec:conclusion}

\modelname{} demonstrates that a conversational model can support both quantitative transcriptomic prediction and language-guided interpretation of experimental samples. Its performance across annotation, perturbation prediction, and biological question answering shows that these capabilities can coexist within a shared model. An established projection interface proves effective in this setting, with the broader contribution lying in how experimental measurements, biological knowledge, and task supervision are brought together during training. GEO-OmicsQA makes interpretation across samples an explicit target for evaluation. These findings support further development of omics-informed language models, while leaving open how training-data composition shapes transfer to new biological problems.

\subsection*{AI use statement}
We used generative AI for synthetic data generation (including mechanistic
hypotheses in training examples), data cleaning and reformatting, method
implementation, and automated answer evaluation.
AI was not used for conceptual frameworks, experiment design, interpretation
of reported results, research hypotheses, or translation; theoretical models,
mathematical claims, and proofs are not applicable.
Additionally, AI assisted coding, manuscript drafting, editing and review,
literature searches, and reference formatting.
Generated data underwent the checks described in the corresponding appendices,
including expert review of a GEO-OmicsQA subset; automated judging assesses
reference agreement, not independent biological validity.
AI-assisted code was reviewed and tested, and authors reviewed manuscript changes.
The authors take responsibility for all content, code, data, and results,
including AI-assisted work.

\bibliographystyle{plainnat}
\bibliography{references}

\appendix

\section{Appendix}

\subsection{Datasets}
\label{app:datasets}

All expression datasets used in this study were aligned to a unified set of 19{,}260 protein-coding and mitochondrial
Ensembl genes derived from GENCODE annotations \citep{mudge2025gencode}. For single-cell datasets, we retained raw count data
prior to normalization, while for bulk expression datasets, we used precomputed transcripts
per million (TPM) values.

\begin{table}[htbp]
  \centering
  \footnotesize
  \caption{Summary of expression datasets used for training and evaluation. Expression profiles are aligned to the same set of 19{,}260 protein-coding and mitochondrial genes. $N_\text{pert}$ denotes the number of distinct genetic perturbations or, for chemical screens, compound--dose conditions; --- indicates an observational study. Pseudobulk denotes scRNA-seq screens aggregated into pseudobulk profiles, which are counted as samples.}
  \label{tab:datasets}
  \setlength{\tabcolsep}{4pt}
  \begin{tabularx}{\linewidth}{@{}>{\raggedright\arraybackslash}X r r r l >{\raggedright\arraybackslash}p{0.14\linewidth}@{}}
    \toprule
    \textbf{Dataset} & \textbf{Samples} & \textbf{Genes} & \boldmath$N_\text{pert}$ & \textbf{Modality} & \textbf{Study type} \\
    \midrule
    \multicolumn{6}{l}{\textit{Observational}} \\
    Human Cell Atlas \citep{regev2017science}  & 6{,}161{,}559 & 19{,}260 & --- & scRNA-seq   & Observational \\
    Tabula Sapiens (GEO) \citep{tabula2022sapiens}   &   483{,}152   & 19{,}260 & --- & scRNA-seq   & Observational \\
    GEO \citep{clough2024geo}  &   615{,}595   & 19{,}260 & --- & Bulk RNA-seq & Observational \\
    GTEx v10 \citep{gtex2020atlas}            &    19{,}616   & 19{,}260 & --- & Bulk RNA-seq & Observational \\
    TCGA \citep{weinstein2013tcga}            &    11{,}504   & 19{,}260 & --- & Bulk RNA-seq & Observational \\
    Human Diseases \citep{schaefer2025cellwhisperer}      &     14{,}112   & 19{,}260 & --- & Bulk RNA-seq & Observational \\
    DepMap Expression 24Q4 \citep{depmap2024}     &     1{,}673   & 19{,}260 & --- & Bulk RNA-seq & Observational \\
    3CA \citep{gavish2023hallmarks} & 4{,}097{,}846 & 19{,}260 & --- & scRNA-seq & Observational \\
    CancerModels.org \citep{perova2022pdcm} & 6{,}437 & 19{,}260 & --- & Bulk RNA-seq & Observational \\
    \midrule
    \multicolumn{6}{l}{\textit{Perturbational}} \\
    DepMap Gene Dependency 24Q4 \citep{depmap2024}      &     1{,}178   & --- & 17{,}916 & CRISPR & Genetic KO \\
    \citet{replogle2022mapping} & 2{,}546{,}823 & 19{,}260 & 9{,}859 & scRNA-seq & Genetic KD \\
    \citet{nadig2025transcriptome}    &   408{,}230   & 19{,}260 &  2{,}390 & scRNA-seq & Genetic KD \\
    Tahoe-100M \citep{zhang2025tahoe} & 46{,}392 & 19{,}260 & 1{,}129 & Pseudobulk & Chemical \\
    sci-Plex3 \citep{srivatsan2020sciplex} & 2{,}259 & 19{,}260 & 752 & Pseudobulk & Chemical \\
    sci-Plex GxE \citep{mcfalinefigueroa2024sciplexgxe} & 15{,}687 & 19{,}260 & 530 & Pseudobulk & Genetic KD + Chemical \\
    \citet{jiang2025pathway} & 1{,}656 & 19{,}260 & 218 & Pseudobulk & Genetic KD \\
    \citet{feng2025crispri} & 8{,}454 & 19{,}260 & 444 & Pseudobulk & Genetic KD \\
    Virtual Cell Challenge H1 hESC \citep{arc2025vcc} & 218{,}268 & 19{,}260 & 150 & scRNA-seq & Genetic KD \\
    X-Atlas/Orion \citep{huang2025xatlas} & 36{,}586 & 19{,}260 & 18{,}310 & Pseudobulk & Genetic KD \\
    \bottomrule{}
  \end{tabularx}
\end{table}

\paragraph{Expression sources.}
The Human Cell Atlas~\citep{regev2017science}, accessed through the CELLxGENE Census
(release 2025-01-30; 27 datasets)~\citep{cellxgene2025census}, is the main source of
single-cell training tasks. The Curated Cancer Cell Atlas (3CA)~\citep{gavish2023hallmarks}
adds 122 human tumour studies with malignant and non-malignant cell annotations. Bulk
profiles come from GTEx v10~\citep{gtex2020atlas}, TCGA~\citep{weinstein2013tcga} with
clinical endpoints from the Pan-Cancer Clinical Data Resource~\citep{liu2018tcgacdr},
DepMap 24Q4 expression and CRISPR dependency data~\citep{depmap2024, pacini2024depmap},
and CancerModels.org~\citep{perova2022pdcm}, which covers patient tumours, xenografts,
organoids, and cell lines. From the Gene Expression Omnibus~\citep{clough2024geo}, we
re-process raw RNA-seq data with the ARCHS4 pipeline~\citep{lachmann2018massive}, which
gives $615{,}595$ human bulk samples. An LLM-based curation pipeline converts their
heterogeneous metadata into sample descriptions and structured annotations (tissue, cell
type, cell line, disease, and perturbation). We split all GEO-derived tasks, including
GEO-OmicsQA, at the publication and study level.

\paragraph{Evaluation-only sources.}
Tabula Sapiens~\citep{tabula2022sapiens} is excluded from training. We use the GEO-hosted
version (GSE201333; 483{,}152 cells, 24 tissues), CellWhisperer reports the same cell count~\citep{schaefer2025cellwhisperer}. The Human Diseases
collection~\citep{schaefer2025cellwhisperer} has $14{,}112$ bulk samples with 27 tissue and
229 disease-subtype labels. We remove its $5{,}683$ samples that overlap with GEO from the
training corpus.

\paragraph{Perturbation and drug-response sources.}
\label{app:pseudobulk_screens}
Genetic screens include the Perturb-seq data of \citet{replogle2022mapping} (K562, RPE1) and
\citet{nadig2025transcriptome} (HepG2, Jurkat), \citet{jiang2025pathway},
\citet{feng2025crispri}, the Virtual Cell Challenge H1 hESC release~\citep{arc2025vcc}, and
X-Atlas/Orion~\citep{huang2025xatlas}. Chemical screens include
Tahoe-100M~\citep{zhang2025tahoe}, sci-Plex3~\citep{srivatsan2020sciplex}, and sci-Plex
GxE~\citep{mcfalinefigueroa2024sciplexgxe}. We aggregate single-cell measurements into
pseudobulk profiles, each matched to an unperturbed control from the same context. Drug
sensitivity labels from GDSC~\citep{yang2013gdsc} and PRISM~\citep{corsello2020prism} are
paired with baseline cell-line profiles. PerturbQA~\citep{wu2025perturbqa} also contributes
direction-of-change and differential-expression training questions. Benchmark (cell line,
perturbed gene) pairs of PerturbQA, and the test pairs of other published perturbation
benchmarks, are removed from all training data.

\paragraph{Knowledge sources.}
Text-only tasks use pathway graphs derived from KEGG~\citep{kanehisa2025kegg} through
BioMaze~\citep{zhao2025biomaze, zhao2025biomaze_graph}, high-confidence human interactions
from STRING v12.0~\citep{szklarczyk2023string}, Gene Ontology and MSigDB gene
sets~\citep{ashburner2000go, geneontology2025, liberzon2015msigdb},
Cellosaurus~\citep{bairoch2018cellosaurus}, UniProt Swiss-Prot~\citep{uniprot2025}, the Cell
Ontology and the disease branch of EFO with Mondo
identifiers~\citep{diehl2016cell, malone2010efo, vasilevsky2022mondo},
DrugMechDB~\citep{gonzalezcavazos2023drugmechdb}, Open Targets
Genetics~\citep{ghoussaini2021otg}, and the training fold of
AssayBench~\citep{debrouwer2026assaybench}. DrugMechDB indications evaluated by the
DrugMechCounterfactuals benchmark are excluded.

\paragraph{Text, question-answering, and general-purpose sources.}
\label{app:instruction_derivatives}
Biomedical question answering and reasoning data come from BioASQ~\citep{krithara2023bioasq},
ReasonMed, Medical-R1, MedReason, and
M23K~\citep{sun2025reasonmed, chen2024huatuogpto1, wu2025medreason, huang2025m1, thapa2025disentangling},
TheBioCollection~\citep{seo2026thebiocollection}, and CAMEL
Biology~\citep{li2023camel}. Relation extraction uses the training splits of
BioRED, BioREx, and ChemProt~\citep{luo2022biored, lai2023biorex, krallinger2017chemprot}. We
also include the profile-linked conversations released with
CellWhisperer~\citep{schaefer2025cellwhisperer}. Scientific text comes from PubMed, PubMed
Central, arXiv, bioRxiv, medRxiv, clinical-trial records, drug labels, and NIH project
abstracts; biomedical code comes from ChemPile-Code~\citep{mirza2025chempile}.
General-purpose data cover encyclopedic text, instruction following, mathematics, coding,
multiple-choice knowledge, dialogue, and tool use (Table~\ref{tab:licenses}). Derived
components do not add data sources: they rephrase existing tasks with output-format
constraints, add follow-up dialogue turns, or add reasoning traces.

\subsection{Cell-Line Detection and Profile Augmentation in Free Text}
\label{app:cell_line_augmentation}
We link cell-line mentions in text to DepMap 24Q4 reference profiles~\citep{depmap2024}
with a dictionary-based detector that uses canonical names, unambiguous Cellosaurus
synonyms~\citep{bairoch2018cellosaurus}, and contextual rules to reject unrelated terms.
After an accepted mention, we insert an \texttt{<omics>} placeholder and keep the original
text: for example, ``MCF-7 cells'' becomes ``MCF-7\texttt{<omics>} cells''. The placeholder
supplies the corresponding profile through the omics projection
(Section~\ref{sec:method_arch}), so that text and profiles are interleaved in one record.

\subsection{LLM-Based Mechanism-of-Action Dataset Generation}
\label{app:moa_dataset}
This dataset links the mechanism of action of each compound to its measured effect on
specific cancer cell lines. For 356 GDSC compounds~\citep{yang2013gdsc}, we first build
compound records from curated drug--target databases, keep per-source provenance, and
assign a three-tier confidence label to each mechanism. Language models are used only where
curated data are missing, and they never overwrite a database fact. For compounds without a
known target, we extract mechanisms from retrieved PubMed articles with three safeguards:
self-consistency across sampling temperatures, deterministic checks of cited PMIDs and gene
symbols, and an adversarial critic. All generative steps use Gemini 3.5
Flash-Lite~\citep{google2026gemini35flashlite}. For each drug--cell-line pair, the generator
then explains the measured response (sensitive or resistant) using only target, pathway, and
mutation facts. Each rationale is checked against an explicit prior: an inhibitor should act
on a pathway that is active at baseline in sensitive cell lines. Hypotheses are labelled
supported or hypothetical, and pairs whose data contradict the prior are excluded. In the
\modelname{} training mixture, these hypotheses serve as rationales for the GDSC
drug-sensitivity tasks. Mechanism paths reserved for the DrugMechCounterfactuals benchmark
are excluded.

\subsection{Marker-Grounded Reasoning Data}
\label{app:markers_cot}
Most annotation tasks supervise only a categorical answer. Our marker-grounded reasoning
data add the expression evidence that supports the answer, for cell type, lineage, immune
identity, cell-cycle phase, tissue, and donor sex across HCA, 3CA, GTEx, TCGA, and GEO
profiles. We follow the evidence-first teacher--student design of scCoT-Synth in
RVQ-Alpha~\citep{li2026rvqalpha}: the teacher (Qwen3.8-27B~\citep{qwen2026qwen38}) sees the
source annotation and an evidence card with detected markers, gene functions, and ontology
context, while the student receives only the question and the expression representation.
Unlike scCoT-Synth, we draw evidence only from external marker panels and published
programmes~\citep{borner2021hra,hu2023cellmarker,karlsson2021hpasc,uhlen2015hpa,palasca2018tissues,uniprot2025,geneontology2025,milacic2024reactome},
keep examples only when the measured markers support the requested level of specificity,
and use grounding checks and separate grading by
Gemma-4-31B-it~\citep{google2026gemma4} to select rationales. Unsupported claims are grounds
for rejection even when the final label is correct.

\subsection{Benchmarks}
\subsubsection{GEO-OmicsQA: Construction and Task Variants}
\label{app:geo_omicsqa_protocol}
GEO-OmicsQA is constructed from GEO studies~\citep{clough2024geo} and their source
publications. Each question refers to one or more real expression profiles, so models must
ground their answers in the linked omics data rather than in textual priors.

\paragraph{Training--test separation.}
We exclude all GEO studies of GEO-OmicsQA test questions, and their source publications,
from the training corpus. This covers expression profiles, metadata, source papers, and all
training examples derived from them. A direct comparison of GEO series, publication, and
sample identifiers found no overlap between the benchmark and the training data.

\paragraph{Task variants.}
Binary QA (1{,}000 questions) is a balanced set of yes/no questions about whether the
profiles support a specific biological claim. Free-text QA (2{,}000 questions) asks for
concise biological explanations and has two equal groups: \textbf{simple} questions without
extra context and \textbf{contextual} questions with auxiliary literature context. Each
family has 500 \textbf{standard} and 500 \textbf{minimal} prompts; minimal prompts reduce
textual scaffolding but keep the same task and answer space. A subset of the test set was
manually reviewed by annotators with biological expertise for question--answer validity,
sample grounding, and biological coherence. Each question--answer pair also includes
supporting passages from the source publication and a short note on how they support the
reference answer. Appendix~\ref{app:geo_omicsqa_examples} shows examples.

\paragraph{Model inputs and prompting.}
Text-only baselines receive ranked gene symbols for each sample, with sample identifiers replaced by dummy labels (\texttt{sample\_1}, \texttt{sample\_2}, \ldots); \modelname{} receives \texttt{<omics>} placeholders with matched embeddings. Models are instructed to answer binary questions with \emph{yes} or \emph{no}; text-only baselines are asked to answer free-text questions in a few sentences. Each example uses one generation request without tools or code execution. Gene lists omit expression magnitudes and may omit genes mentioned in the question, even with the larger budget.

\paragraph{Dynamic gene budgets.}
We cap each profile at 1{,}000 genes, following the setting illustrated in the C2S-Scale model card.\footnote{\url{https://huggingface.co/vandijklab/C2S-Scale-Gemma-2-27B}} For multi-sample prompts, we reduce the per-profile count as needed to fit the context window, consistent with the C2S authors' multi-cell formatting procedure \citep{rizvi2025scaling}.
For a prompt containing $P$ distinct profiles, we initially allocate
\[
 n = \min\!\left(1000,\left\lfloor\frac{L-H}{P}\right\rfloor\right)
\]
genes per profile, where $L$ is the model's input or context limit and $H$ is the reserved headroom. Because a gene symbol can span several tokens, we then count the complete formatted prompt and shorten all gene prefixes equally until it fits within $L-H$. Counting includes the question, instructions, and chat template where applicable, using provider token-count APIs or the local tokenizer. No question or profile is dropped. Repeated sample identifiers count as one profile. Each of the nine baselines is evaluated on all six 500-example splits.

\begin{table}[htbp]
\centering
\small
\caption{Selected GEO-OmicsQA input budgets. Gemini limits are input limits; the GPT input limit is separate from its total context. Headroom is the minimum reserved amount, not the actual unused capacity.}
\label{tab:geo_qa_budgets}
\begin{tabular}{lrrr}
\toprule
Model & Input/context limit & Headroom & Genes/profile \\
\midrule
Gemini Flash / Pro & 1{,}048{,}576 & 1{,}000 & 1{,}000 \\
GPT Sol / Luna & 922{,}000 & 10{,}000 & 1{,}000 \\
Qwen 9B / 27B & 262{,}144 & 4{,}096 & 1{,}000 \\
Cell-o1 & 32{,}768 & 4{,}096 & 330--1{,}000 \\
C2S-Scale-27B & 8{,}192 & 1{,}000 & 80--1{,}000 \\
\bottomrule
\end{tabular}
\end{table}

\paragraph{Generation and scoring settings.}
Gemini uses temperature 0 and a 10{,}000-token output cap; GPT retains API-default sampling and the same cap. API reasoning controls are not explicitly overridden. Local models use vLLM with temperature 0; Qwen and Cell-o1 use chat templates with thinking disabled and a 4{,}096-token output cap. C2S uses its raw prompt format with a 1{,}000-token free-text cap. For C2S binary questions, we select the higher-probability first-token continuation among \texttt{Yes} and \texttt{No}; other baselines generate their answers. Binary subsets are class-balanced, so accuracy equals balanced accuracy. Each result is from one run; the comparisons do not control for training exposure or reasoning budget. Prompts are checked against their input budgets. Two Pro, 11 Cell-o1, 10 Qwen 9B, and four Qwen 27B responses reached their output caps and remain included in scoring. Judge scores measure agreement with the literature-derived references, not an independent audit of every claim against the supplied profiles.

\paragraph{C2S-Scale output behavior.}
\label{app:c2s_geoqa_behavior}
We evaluate the released \path{vandijklab/C2S-Scale-Gemma-2-27B} checkpoint without additional QA fine-tuning. Its poor free-text result primarily reflects an output-format failure: it often continues with a list of gene symbols instead of answering the biological question. For example, the first standard simple question asks which of two sorted populations has an arterial versus hematopoietic signature; the response begins \texttt{10000001156F18RIK RPL37A RPL37 RPS27 RPL32 RPL39 RPL36 RPS29 RPL41 RPS18} and continues as a gene list. In an audit of all standard simple outputs, 60\% of them have at least 80\% of their word tokens matching the expression dataset's gene-symbol vocabulary; the median matching fraction is 92.3\%. These are output-format diagnostics, not biological correctness measures.

The C2S authors distinguish their released pretrained checkpoints from downstream QA-finetuned models and report that cell-generation training can make the released models produce gene sentences for many prompts.\footnote{Author clarifications: \url{https://github.com/vandijklab/cell2sentence/issues/15\#issuecomment-3397785415} and \url{https://github.com/vandijklab/cell2sentence/issues/16\#issuecomment-3407399888}.} Our result therefore characterizes the released checkpoint under this zero-shot protocol, not the performance of a QA-finetuned C2S model. The binary logit comparison forces a label even when neither word is the preferred natural continuation: it selects \emph{Yes} for 959 of 1{,}000 questions, with 98.0\% Yes recall and 6.2\% No recall, yielding 52.1\% balanced accuracy.

\paragraph{Sensitivity to expression profiles.}
Correct profiles yield 76.7\% binary accuracy and a mean free-text judge score of 0.65105, compared with 69.5\% and 0.61445 for incorrect profiles sampled from the lowest-cosine 5\% of eligible profiles. The moderate drop in performance suggests that expression profiles contribute to answering GEO-OmicsQA questions.

\subsubsection{Constrained Likelihood Scoring}
\label{app:constrained_likelihood}
For closed-set classification tasks, each candidate label is treated as a possible assistant continuation. We encode the prompt once, score each label token sequence under teacher forcing, sum token log probabilities, and normalize by label length. The top-scoring label gives accuracy. To compute macro AUROC, we apply a softmax over the candidate label scores for each example, compute one-vs-rest AUROC for each class with both positive and negative examples, and average over these classes. We report these likelihood-based AUROC values where prior work uses classifier-style metrics, while retaining model-specific continuation formatting.

\paragraph{Ranked-gene annotation baselines.}
\label{app:annotation_baselines}
C2S wording follows the released cell-type and tissue templates of
\citet{rizvi2025scaling}, with a fixed adaptation for disease subtype.
Candidate continuations use a newline, lowercase label, and period for C2S,
and bare lowercase labels for Cell-o1. C2S runs use the default
Transformers SDPA backend, which omits Gemma-2 attention-logit soft-capping.

\subsubsection{DeepEval Judge for Open-Ended Generation}
\label{app:deepeval_judge}
Open-ended generation tasks, including GEO-OmicsQA simple and contextual tasks, are scored with
DeepEval GEval \citep{liu2023geval} using the Gemini judge model
\texttt{gemini-3.1-flash-lite} \citep{google2026gemini31flashlite}. Each judged test case is passed to GEval
with three fields: the input question, the model's actual output, and the expected
reference output. The reported judge score is the mean DeepEval score.

The judge follows six evaluation steps: it reads the question; identifies the key biological entities in the expected output (genes, cell types or lines, tissues, organisms, diseases, treatments, compounds, and pathways); checks whether the actual output mentions the same entities or accepted synonyms, without penalizing additional correct details; verifies that directional claims agree; evaluates whether the biological interpretation and mechanistic reasoning are consistent with the expected output; and treats a missing central finding as a significant deficiency. The rubric assigns 0--2 to irrelevant, unanswered, or wrong-subject answers; 3--4 to the correct topic with major factual errors; 5--6 to partially correct but incomplete or inaccurate answers; 7--8 when the main finding and most entities are correct; and 9--10 when all key entities, directions, and conclusions agree.

\subsubsection{PerturbQA: Protocol and Detailed Results}
\label{app:perturbation_prediction}

\paragraph{Test population and split scope.}
We use the released \texttt{split=test} rows in the four \texttt{<cell>-dir.csv} files at PerturbQA revision \texttt{df2fe3a7802f2a46a43876b4f8674734aad3a30a}. Each example is identified by (cell line, perturbed gene, readout gene). Canonicalized identities and gold labels are checked against the official files; duplicate, missing, and extra rows are rejected. There is no new random split. K562 contributes 3{,}119 rows and 150 readout genes eligible for AUROC; RPE1, 5{,}365 and 193; HepG2, 3{,}603 and 166; and Jurkat, 4{,}121 and 181. We do not evaluate the additional K562-Set dataset or the differential-expression and gene-set-enrichment tasks here.

The data construction policy excludes benchmark (cell line, perturbed gene) pairs from genetic training screens in the same cell line, while allowing the same gene in other cell lines. We apply the same cell-line/perturbation holdout policy to all other training templates derived from the source data used to construct the PerturbQA test set, ensuring that held-out pairs cannot enter training through alternative task formulations. This is a cell/perturbation holdout policy, not a globally unseen-gene split. The present scoring audit verifies test alignment, not the complete training exposure of the evaluated checkpoints. The scoring audit does not establish leakage-free generalization across training records, extended screens, and retrieved pair notes.

\paragraph{Model inputs and inference.}
Each prompt asks whether knocking down a specified gene increases or decreases the expression of a readout gene in a given cell line, requesting a single-word answer. Inputs also include public notes on the gene pair and the cell line's unperturbed 20{,}541-dimensional control profile, supplied through the omics projection. We sample three answers per question with temperature 0.6, top-$p$ 0.9, no top-$k$ restriction, an 8{,}192-token context, and a 4{,}096-token output cap. Each question/replicate receives a distinct reproducible SHA-256-derived seed; prompt-token hashes agree across runs. All 48{,}624 outputs were valid and none were truncated.

\paragraph{Three-run answer aggregation.}
Following SUMMER's methodology~\citep{wu2025perturbqa}, we aggregate three sampled answers per question before computing AUROC to enable direct comparison.
For each question, \modelname{} maps Increase/Up to 1 and Decrease/Down to 0 and averages the three sampled answer scores. Invalid or truncated answers are excluded from this mean; if all three fail, the fallback is the mean label for that cell/readout gene in official TRAIN, never TEST. This fallback was not needed. AROMA retains its native three-option prompt and maps Up to 1, unchanged (ND) to 0.5, and Down to 0; invalid or truncated answers also receive 0.5. Its score is the arithmetic mean of all three values. We average answers before AUROC, not the three run-level AUROCs, and do not select a best run. Every run is checked against the same complete official test identities and labels.

\paragraph{AUROC.}
Up is the positive class. Let $G_c$ be the readout genes with both gold classes in cell line $c$. We compute
\[
  A_c=\frac{1}{|G_c|}\sum_{g\in G_c}
  \operatorname{AUROC}\!\left(\{(y_{cpg},s_{cpg}):p\text{ in TEST}\}\right),
  \qquad A=\frac{1}{4}\sum_c A_c.
\]
Groups with only one gold class are omitted from AUROC, but not from F1. The implementation checks agreement with the pinned official evaluator. \modelname{} parsing accepts only completed bare labels or final answers following closed thinking blocks. With three valid binary answers, the averaged score takes values 0, 1/3, 2/3, or 1, providing a ranking beyond a single binary answer. Released baseline scores are preserved rather than converted to hard labels for AUROC.

\paragraph{Direction F1.}
We pool all direction examples within each cell and calculate $F_{1,c,k}=2\mathrm{TP}_{c,k}/(2\mathrm{TP}_{c,k}+\mathrm{FP}_{c,k}+\mathrm{FN}_{c,k})$ for $k\in\{\mathrm{Up},\mathrm{Down}\}$. Class F1 is averaged equally across the four cells; macro-F1 is the mean of the two class averages. For \modelname{} and AROMA, the averaged score predicts Up above 0.5 and Down below 0.5; an exact tie predicts neither direction and remains a false negative for its gold class. No test question is removed because of a failed prediction. This population contains no gold unchanged examples, so it cannot measure false positives on unchanged genes or substitute for three-class F1.

\paragraph{Baseline provenance and decision rules.}
SUMMER, retrieval without LLM, LLM without retrieval, LLM without chain-of-thought, GEARS, GenePT, GAT, and MLP are rescored from the authors' \texttt{results.zip} release~\citep{wu2025perturbqa,wu2025perturbqadata}; they are not retrained locally. The archive SHA-256 is \nolinkurl{d6d3f3861043e5e0f02418c9213ca20f4167dc82bd96ff2ed5af23970ff59df0}. SUMMER's released scores already aggregate sampled answers and are not averaged again. SUMMER and the other scalar-score baselines retain their released \texttt{dir\_pred} for AUROC. Their diagnostic F1 predicts Up above 0.5 and Down below 0.5; exact ties predict neither. This does not establish the authors' intended hard-decision calibration, especially for GEARS, or recover SUMMER's individual abstentions. For GAT and MLP, AUROC uses the Up logit minus the Down logit, a diagnostic conversion whose equivalence to the original paper is unverified. Their F1 uses native three-class argmax (ND, Down, Up); ND predicts neither direction. GenePT metrics are computed separately for each of three released replicates, then averaged within cells before averaging cells. No replicate is selected by TEST performance. Matched direction predictions for scGPT and STATE are unavailable, so their published three-class scores are not imported into these tables.

\begin{table}[t]
\centering
\caption{\textbf{Complete PerturbQA direction comparison.} All values are on a 0--1 scale and averaged equally over cells. $\dagger$: diagnostic threshold-based F1; $\ddagger$: diagnostic logit-to-AUROC conversion. These F1 scores are newly computed on the direction subset, not published three-class scores. SynthPert (paper) is a published reference, not a matched reproduction; --- denotes unavailable F1.}
\label{tab:perturbqa_all}
\small
\begin{tabular}{lrrrr}
\toprule
Model & AUROC & F1 Up & F1 Down & Macro-F1 \\
\midrule
\modelname{}-9B & 0.7635 & 0.8772 & 0.8749 & 0.8761 \\
SUMMER & 0.6547 & 0.7977$^{\dagger}$ & 0.7908$^{\dagger}$ & 0.7942$^{\dagger}$ \\
Retrieval without LLM & 0.6431 & 0.7970$^{\dagger}$ & 0.7779$^{\dagger}$ & 0.7875$^{\dagger}$ \\
GenePT protein & 0.5746 & 0.7590$^{\dagger}$ & 0.7304$^{\dagger}$ & 0.7447$^{\dagger}$ \\
GenePT gene & 0.5681 & 0.7288$^{\dagger}$ & 0.6968$^{\dagger}$ & 0.7128$^{\dagger}$ \\
GAT & 0.6043$^{\ddagger}$ & 0.6963 & 0.6778 & 0.6871 \\
MLP & 0.4976$^{\ddagger}$ & 0.6451 & 0.6348 & 0.6400 \\
LLM without retrieval & 0.5224 & 0.5401$^{\dagger}$ & 0.6061$^{\dagger}$ & 0.5731$^{\dagger}$ \\
GEARS & 0.5775 & 0.3411$^{\dagger}$ & 0.5885$^{\dagger}$ & 0.4648$^{\dagger}$ \\
LLM without CoT & 0.4970 & 0.0994$^{\dagger}$ & 0.6354$^{\dagger}$ & 0.3674$^{\dagger}$ \\
AROMA & 0.5788 & 0.7064 & 0.6803 & 0.6934 \\
\midrule
SynthPert (paper)~\citep{phillips2025synthpert} & 0.6875 & --- & --- & --- \\
\bottomrule
\end{tabular}
\end{table}

\begin{table}[t]
\centering
\caption{\textbf{Per-cell PerturbQA results.} Each entry is AUROC / direction macro-F1. Diagnostic conversions and threshold caveats are as in Table~\ref{tab:perturbqa_all}. GenePT entries average replicate metrics. SynthPert values are published AUROCs with unavailable direction F1.}
\label{tab:perturbqa_cells}
\footnotesize
\setlength{\tabcolsep}{3pt}
\begin{tabular}{lrrrr}
\toprule
Model & K562 & RPE1 & HepG2 & Jurkat \\
\midrule
\modelname{}-9B & 0.7939 / 0.8482 & 0.7173 / 0.8907 & 0.7679 / 0.8891 & 0.7750 / 0.8763 \\
SUMMER & 0.6379 / 0.7143 & 0.6426 / 0.8422 & 0.6634 / 0.8311 & 0.6748 / 0.7894 \\
Retrieval without LLM & 0.6913 / 0.7164 & 0.6331 / 0.8478 & 0.6380 / 0.8239 & 0.6098 / 0.7618 \\
GenePT protein & 0.5583 / 0.6471 & 0.5857 / 0.8224 & 0.5558 / 0.7846 & 0.5986 / 0.7246 \\
GenePT gene & 0.4891 / 0.5871 & 0.5891 / 0.7994 & 0.6112 / 0.7751 & 0.5828 / 0.6896 \\
GAT & 0.5949 / 0.5381 & 0.6054 / 0.7885 & 0.6429 / 0.7613 & 0.5742 / 0.6603 \\
MLP & 0.5223 / 0.4828 & 0.5136 / 0.7473 & 0.4953 / 0.7127 & 0.4594 / 0.6171 \\
LLM without retrieval & 0.4984 / 0.4916 & 0.5223 / 0.6232 & 0.5311 / 0.5604 & 0.5376 / 0.6172 \\
GEARS & 0.6660 / 0.3433 & 0.6064 / 0.5941 & 0.5195 / 0.4553 & 0.5181 / 0.4665 \\
LLM without CoT & 0.5019 / 0.2063 & 0.4919 / 0.3990 & 0.4898 / 0.4065 & 0.5044 / 0.4578 \\
AROMA & 0.5645 / 0.6069 & 0.5683 / 0.7429 & 0.5987 / 0.7256 & 0.5838 / 0.6981 \\
\midrule
SynthPert (paper)~\citep{phillips2025synthpert} & 0.65 / --- & 0.73 / --- & 0.72 / --- & 0.65 / --- \\
\bottomrule
\end{tabular}
\end{table}

\paragraph{SynthPert: published reference only.}
SynthPert's Table~2 reports direction AUROCs of 0.65, 0.73, 0.72, and 0.65 for K562, RPE1, HepG2, and Jurkat, respectively~\citep{phillips2025synthpert}; their arithmetic mean is 0.6875. We cite the specific arXiv v1 table from which these rounded values were taken. Its trained checkpoint is not publicly available, so we report the published results rather than evaluating the model on our matched test set. The tables label them as published references, separate from locally scored results; matched AUROC and direction F1 are unavailable. Its three-class formulation predicts Up, Down, and not differentially expressed jointly, unlike the conditional direction task used here.

\paragraph{AROMA reproduction.}
We generated AROMA predictions from the authors' released multimodal checkpoint and augmented three-class prompts~\citep{wang2026aroma}, preserving the original chat template and protein and graph feature placeholders. The current three-run evaluation generates predictions only for the 16{,}208 official direction TEST questions; it does not rerun the full three-class test. Feature lookup preserves case-sensitive gene names; canonicalized identifiers are used separately for test matching. Checkpoint tensors, active adapters, feature mapping, prompt formatting, multimodal injection, and output extraction were audited.

The answer mapping and aggregation follow the three-run protocol above. AROMA obtains 0.5788 mean direction AUROC and 0.6934 direction macro-F1 (Tables~\ref{tab:perturbqa_all} and~\ref{tab:perturbqa_cells}). Inference retains the released sampling settings: temperature 0.7, top-$k$ 20, top-$p$ 0.95, and 512 new tokens, with batch size 32 per GPU and two GPU workers per run. Each replicate/worker uses a distinct SHA-256-derived PyTorch seed and an advancing random stream. These decoding settings differ from \modelname{}'s; only the test population and score aggregation are matched. All 48{,}624 predictions are retained, including 108 invalid or truncated answers scored as 0.5. Runtime package versions differ from the authors' requirements, so this is an independent reproduction, not a bitwise replication.

As a reproduction check, an earlier single-run evaluation on all 116{,}041 three-class TEST rows gave AROMA a macro-F1 of 0.7300, compared with 0.7325 from the rounded per-cell values in its paper~\citep{wang2026aroma}.

\subsection{Training Details}
\label{app:training_details}

We provide optimization, sampling, and supervision details for the three training stages introduced in Section~\ref{sec:training_procedure}.

\subsubsection{Stage 1: Multimodal Pretraining}
\label{app:pretraining}
This stage jointly performs continued pretraining (CPT) on free-text records and supervised fine-tuning (SFT) on instruction examples, interleaving both throughout training.

\paragraph{Initialization and optimization.}
We initialize the language backbone from the text-model weights of the official pretrained Qwen3.5-9B checkpoint and augment it with the omics projection and gene vocabulary described in Section~\ref{sec:method_arch}. The language backbone, token embeddings, and omics projection are updated jointly; the omics encoders remain frozen and their outputs are precomputed. We train for 40{,}700 optimizer updates on 48 NVIDIA H100 80GB GPUs. Each GPU processes one microbatch of up to 10{,}240 input tokens per update, with gradient accumulation set to one, giving a nominal global batch size of 491{,}520 input tokens. The measured total is 19.999 billion input tokens. Token counts include masked context and omics input positions and exclude padding. We use 8-bit AdamW with $\beta_1=0.9$, $\beta_2=0.95$, $\epsilon=10^{-8}$, weight decay $0.1$, and gradient-norm clipping at $1.0$. The learning rate increases linearly to $2\times10^{-5}$ over 200 updates, then follows a cosine schedule with a minimum of $2\times10^{-6}$. We use BF16 model weights and gradients, gradient checkpointing, and distributed data parallelism. The recorded training runtime was approximately 43.8 hours, corresponding to 2{,}103 H100 GPU-hours across 48 GPUs.

\paragraph{Source sampling.}
We assign a sampling weight to each data subset using source-size estimates and manual adjustments to balance the number of examples, token volume, and coverage of different task types. These weights define target shares of input tokens, including prompt tokens masked from the loss. Table~\ref{tab:pretraining_mixture} reports the observed input-token share for the completed run using the categories defined in Section~\ref{sec:datasets}.

\begin{table}[htbp]
    \centering
    \small
    \caption{Observed Stage 1 input-token shares by training-data category. Counts include masked context and repeated exposure, and exclude padding. Human genetics was not included in this stage. Values are rounded.}
    \label{tab:pretraining_mixture}
    \begin{tabular}{@{}lr@{}}
        \toprule
        \textbf{Training-data category} & \textbf{Input-token share (\%)} \\
        \midrule
        Cellular characterization & 12.96 \\
        Expression interpretation and reasoning & 18.93 \\
        Perturbation response & 7.94 \\
        Gene essentiality and functional screens & 2.92 \\
        Drug response and mechanisms & 1.06 \\
        Biological knowledge and networks & 3.25 \\
        Biomedical QA and information extraction & 5.45 \\
        Scientific and biomedical text and code & 40.62 \\
        General-purpose capabilities & 6.88 \\
        \midrule
        Total & 100.00 \\
        \bottomrule
    \end{tabular}
\end{table}

\paragraph{Sequence construction and loss.}
The maximum sequence length is 10{,}240 tokens. Longer free-text records are divided into consecutive windows with boundary overlap for next-token prediction. For an overlength conversation, we sample a window containing supervised tokens. Window boundaries preserve complete omics spans. Windows are packed into token-budgeted batches while keeping sequences independent. The first token of each window, omics placeholders, and their delimiters are excluded from the loss. Free-text records otherwise receive next-token supervision throughout, while conversational records receive supervision on assistant responses. The objective is averaged over supervised tokens across devices and accumulation steps.

\subsubsection{Stage 2: Supervised Instruction Tuning}
\label{app:sft_training}

\paragraph{Initialization and optimization.}
We initialize Stage 2 from a Stage 1 checkpoint. The language backbone, token embeddings (including the added gene tokens), and omics projection continue to be updated jointly. The omics encoders remain frozen, and their precomputed outputs are supplied through the same projection as in Stage 1. We use AdamW with $\beta_1=0.9$, $\beta_2=0.999$, weight decay $0.01$, and gradient-norm clipping at $1.0$. The learning rate increases linearly to $1.4\times10^{-5}$ over 169 updates (5\% of the training budget, rounded down), then follows a cosine schedule to $1.4\times10^{-6}$. Training uses BF16 computation with FP32 master weights and gradient reduction, gradient checkpointing, and hybrid sharded data parallelism: parameters are sharded across the eight GPUs within each node and replicated across six nodes.

\paragraph{Batching and training budget.}
We train for two epochs on 48 NVIDIA H100 80GB GPUs, with a global batch of 3{,}072 conversations per optimizer update (64 per GPU). Each local batch is divided into dynamic microbatches with a budget of 32{,}768 input tokens per GPU; gradients accumulate over these microbatches before the optimizer update. The number of microbatches depends on sequence lengths. We use padding-free packing with independent attention and recurrent state for each conversation, and a maximum sequence length of 32{,}768 tokens. All training conversations fit within this limit, so no truncation is required. Dropping the final incomplete batch gives 1{,}696 updates per epoch and 3{,}392 updates in total, with 2.788 billion input tokens processed across the two epochs. Token counts include masked context and omics input positions and exclude padding. The training job ran for approximately 12.8 hours, corresponding to 615 H100 GPU-hours, including initialization, validation, and checkpoint writing. The model after the second epoch initializes reinforcement learning.

\paragraph{Training mixture.}
The corpus contains 5{,}210{,}786 training conversations drawn from 104 data components, combining expression-conditioned biological tasks with text-only biological knowledge and general-purpose instructions. Of these conversations, 3{,}489{,}641 contain omics inputs and 864{,}886 contain nonempty reasoning traces; these subsets overlap. The biological examples cover cellular characterization, expression interpretation, perturbation responses, gene essentiality, functional screens, drug sensitivity and mechanisms, human genetics, and biological knowledge and reasoning. Mathematics, coding, general knowledge, instruction following, and tool-use conversations complement the biological examples. The mixture includes direct answers, reasoning followed by a final answer, and multi-turn interactions. Counts refer to conversation records before repeated exposure across epochs; multiple tasks or instruction variants can share the same underlying biological sample. Table~\ref{tab:task_examples} summarizes the corpus as 157 task groups by biological objective and input/output format, with representative examples.

\paragraph{Conversation formatting and supervision.}
We serialize conversations using a training adaptation of the backbone's chat template that preserves reasoning across assistant turns, message order, and omics references. We minimize next-token negative log-likelihood only on assistant turns selected for supervision. System, user, and tool messages, together with explicitly excluded assistant turns, remain available as context but do not contribute to the loss. The corpus contains 3{,}437 such excluded assistant turns in agent trajectories. On supervised turns, supplied reasoning traces and final answers are both training targets; synthetic empty reasoning prefixes inserted by the formatter are masked. This supports both direct-answer and reasoning examples without making an empty rationale a supervised target.

\subsubsection{Stage 3: Reinforcement Learning}
\label{app:rl_training}

\paragraph{Policy initialization.}
We initialize GRPO from the multimodal Qwen3.5-9B model after two epochs of supervised instruction tuning on a curated corpus of 5{,}210{,}786 conversations. SFT starts from an intermediate model obtained during multimodal continued pretraining. The resulting SFT policy provides both the initial actor and the frozen reference policy for RL. GRPO follows SFT directly, without an intervening preference-learning stage. The language backbone, gene-token embeddings, and omics projection remain trainable, while the pretrained omics encoders remain frozen.

\paragraph{Reasoning initialization and response format.}
We use the preceding CPT and SFT stages as the RL warm start, without an additional RL-specific cold-start SFT run. Multimodal CPT develops the connection between expression profiles, gene tokens, and language. SFT then supplies both direct answers and supervised reasoning: the SFT corpus contains 864{,}886 conversations with nonempty reasoning traces, alongside instruction, mathematical, biological, and tool-use examples. The SFT formatter preserves reasoning on supervised assistant turns, applies loss only to the selected assistant tokens, and masks synthetic empty reasoning prefixes. System, user, tool, and excluded assistant turns remain context. Consequently, direct-answer examples are not converted into targets that explicitly teach an empty rationale.

RL uses the native thinking-enabled chat template. The prompt supplies the opening \texttt{<think>} tag, and a valid continuation supplies exactly one \texttt{</think>} followed by a nonempty final answer. The reward adapter restores the opening tag for parsing when it is absent from the generated continuation. Missing answer boundaries, additional chat-role delimiters, or deprecated \texttt{<answer>} wrappers receive zero reward. An empty reasoning block is structurally valid but earns no evidence-grounding credit. The verifier scores the final answer separately from the rationale; there is no reward for merely producing a longer chain of thought. These are the RL training settings, not a claim that every benchmark is decoded with thinking enabled.

\paragraph{Exploratory comparison of post-training strategies.}
Before training the 9B model, we compared SFT, direct GRPO, Kahneman--Tversky Optimization (KTO; \citealp{ethayarajh2024kto}), Noise Contrastive Alignment (NCA; \citealp{chen2024noise}), and preference learning followed by GRPO using Qwen3~1.7B backbones (Table~\ref{tab:rl_small_model_motivation}). KTO and NCA are alternative alignment methods applied after SFT. We evaluate biological performance on the annotation and GEO-OmicsQA tasks introduced in Sections~\ref{sec:results_zero_shot_annotation} and~\ref{sec:results_geo_qa}. MMLU multiple-choice knowledge questions and a GSM8K mathematical word-problem probe serve as diagnostics of general-capability changes relative to these SFT policies.

Direct GRPO from SFT epoch~2 yields higher observed scores on all eight reported metrics relative to that initializer. GEO-OmicsQA binary accuracy rises from 74.0 to 74.4 and the free-text score from 55.5 to 57.4, while the four annotation AUROCs increase by 0.1--0.4 percentage points. MMLU rises from 34.29 to 54.37 and GSM8K from 72.50 to 75.75. Direct GRPO achieves the highest GEO-OmicsQA scores among the compared workflows, although some preference-learning alternatives achieve higher annotation or general-capability scores. No workflow leads on every metric. For the final 9B model, we use multimodal pretraining followed by two SFT epochs and direct GRPO, without a KTO or NCA stage.

These are descriptive comparisons from single training runs with separately filtered prompt sets and unequal budgets. Direct GRPO starts from SFT epoch~2, whereas KTO and NCA start from SFT epoch~1. The 1.7B models also have a different pretraining lineage from the final 9B model. Consequently, this table does not isolate the effect of the alignment method or establish general-capability retention for the final 9B model or relative to the original pretrained backbone.

\begin{table}[htbp]
    \centering
    \small
    \setlength{\tabcolsep}{4pt}
    \caption{Exploratory Qwen3~1.7B comparison of biological performance and general capabilities. All values are on a 0--100 scale; higher is better. Tabula Sapiens and Human Diseases report macro AUROC; GEO-OmicsQA reports aggregate binary accuracy and free-text judge score. MMLU and the GSM8K probe use thinking-mode decoding and track general-capability changes relative to SFT. KTO and NCA start from SFT epoch~1. The KTO$\rightarrow$GRPO result is an intermediate 100-update checkpoint. Each workflow is a single training run.}
    \label{tab:rl_small_model_motivation}
    \resizebox{\linewidth}{!}{%
    \begin{tabular}{l*{8}{r}}
        \toprule
        & \multicolumn{2}{c}{Tabula Sapiens} & \multicolumn{2}{c}{Human Diseases} & \multicolumn{2}{c}{GEO-OmicsQA} & \multicolumn{2}{c}{General capabilities} \\
        \cmidrule(lr){2-3}\cmidrule(lr){4-5}\cmidrule(lr){6-7}\cmidrule(lr){8-9}
        Qwen3~1.7B workflow & Cell type top-20 & Tissue & Disease subtype & Tissue & Binary QA & Free text & MMLU & GSM8K \\
        \midrule
        SFT epoch 1 & 93.2 & 82.1 & 83.5 & 96.1 & 73.1 & 54.6 & 34.96 & 71.25 \\
        SFT epoch 2 & 94.4 & 81.9 & 83.2 & 96.7 & 74.0 & 55.5 & 34.29 & 72.50 \\
        SFT epoch 2 $\rightarrow$ GRPO & 94.8 & 82.3 & 83.6 & 96.8 & 74.4 & 57.4 & 54.37 & 75.75 \\
        KTO & 94.2 & 82.3 & 82.9 & 96.6 & 71.6 & 55.7 & 47.78 & 73.50 \\
        NCA & 94.9 & 82.9 & 83.5 & 96.9 & 73.5 & 56.7 & 45.69 & 74.50 \\
        NCA $\rightarrow$ GRPO & 95.1 & 83.0 & 83.7 & 96.8 & 73.0 & 56.5 & 57.60 & 76.25 \\
        KTO $\rightarrow$ GRPO (100) & 94.5 & 82.5 & 83.4 & 96.6 & 72.7 & 56.1 & 37.54 & 74.25 \\
        \bottomrule
    \end{tabular}}
\end{table}

\paragraph{RL sources and task coverage.}
Table~\ref{tab:rl_sources_9b} summarizes the 19 reward-task families and their prompt counts. Biological classification and marker-reasoning tasks draw on the Human Cell Atlas (HCA), Curated Cancer Cell Atlas (3CA), GTEx, TCGA, GEO, and CancerModels.org datasets. Gene-list tasks use expression-derived and curated gene sets. Gene-essentiality tasks use DepMap, while drug-sensitivity tasks use GDSC and the PRISM primary and secondary activity screens. Differential-expression and direction tasks use PerturbQA~\citep{wu2025perturbqa}, and the four mechanism-of-action tasks use DrugMechDB~\citep{gonzalezcavazos2023drugmechdb}. Graph-path questions use a biological graph assembled from STRING resources.

To retain general capabilities, the candidate pool also includes GSM8K training questions~\citep{cobbe2021gsm8k}, multiple-choice questions from MMLU auxiliary training and ARC training splits~\citep{hendrycks2020mmlu,clark2018arc}, and CAMEL biology questions augmented with programmatically checkable instructions~\citep{li2023camel}. UltraData-RL-2609 contributes only its Math and Knowledge configurations~\citep{openbmb2026ultradatarl}; its Code and Long-Context configurations are not used. Agent/tool-use SFT data, Open Targets Genetics, AssayBench, and the UltraInteract preference data should not be conflated with this RL mixture. RL records contain prompts, optional omics inputs, reference answers, and verifier metadata; optimization uses newly sampled responses rather than treating source rationales or filtering completions as supervised targets.

\begin{table}[htbp]
    \centering
    \small
    \setlength{\tabcolsep}{5pt}
    \caption{Audited 9B RL candidate pool and training prompts selected using the SFT initializer. Counts refer to prompts, not the eight sampled completions per prompt.}
    \label{tab:rl_sources_9b}
    \begin{tabular}{lrr}
        \toprule
        Task / source & Candidates & Selected \\
        \midrule
        GDSC drug sensitivity & 4{,}779 & 2{,}853 \\
        DepMap gene essentiality & 4{,}227 & 565 \\
        PerturbQA: differential expression & 2{,}510 & 334 \\
        PerturbQA: direction & 2{,}503 & 771 \\
        PRISM primary activity & 1{,}891 & 1{,}202 \\
        PRISM secondary activity & 1{,}904 & 1{,}288 \\
        Biological label classification & 3{,}479 & 998 \\
        Marker reasoning & 3{,}800 & 60 \\
        Gene-list / gene-set prediction & 3{,}001 & 14 \\
        Biological graph paths & 5{,}679 & 557 \\
        DrugMechDB: closed edge & 607 & 245 \\
        DrugMechDB: open edge & 245 & 155 \\
        DrugMechDB: effect & 431 & 287 \\
        DrugMechDB: masked node & 620 & 163 \\
        GSM8K mathematics & 2{,}815 & 917 \\
        MMLU / ARC multiple choice & 1{,}892 & 402 \\
        Checkable instruction following & 2{,}868 & 1{,}314 \\
        UltraData-RL-2609: Math & 1{,}999 & 26 \\
        UltraData-RL-2609: Knowledge & 1{,}000 & 20 \\
        \midrule
        Total & 46{,}250 & 12{,}171 \\
        \bottomrule
    \end{tabular}
\end{table}

\paragraph{Data audit.}
Before difficulty filtering, we remove superseded, inconsistent, and evaluation-reserved prompts, and prompts that fail overlap, duplicate, or semantic-identity checks against protected splits. This gives the audited pool of 46{,}250 prompts. The audit also checks that reference answers are compatible with the task verifiers and that all referenced expression vectors are finite and have the expected 20{,}541 dimensions. Prompts allow at most eight profiles and 1{,}024 tokens. This audit is defined by registered identity keys; it is not a claim of universal benchmark decontamination, and some RL prompts appeared during SFT.

\paragraph{Model-based difficulty filtering.}
Using the SFT initializer, we sample $G=8$ responses per prompt with temperature $1.2$ and a 2{,}048-token response limit. Let $c_i\in\{0,1\}$ denote task-verifier correctness after structural and rejection gates. We retain a prompt if
\begin{equation}
    1 \leq \sum_{i=1}^{8} c_i \leq 6.
    \label{eq:rl_selection}
\end{equation}
The lower bound retains one-success groups, which are difficult but demonstrably solvable within the sampling budget. Zero-success groups provide no observed correct trajectory, while seven- and eight-success groups are removed to concentrate training on less-saturated problems. Selection uses binary correctness, not the shaped scalar reward; otherwise variation in formatting or rationale length could make an outcome-saturated prompt appear useful. The selected prompt set is frozen before RL, with fresh responses sampled during every optimization step.

Difficulty filtering covers 35{,}609 of the 46{,}250 candidate prompts (77.0\%; Table~\ref{tab:rl_filter_funnel}). Only complete groups of eight sampled responses are considered for selection; incomplete groups and unsampled prompts are excluded. A response that reaches the generation limit still belongs to its completed sampling group, but receives zero if it never produces the required final-answer boundary. The selected task mixture therefore reflects both sampling coverage and verifier correctness. We uniformly shuffle the selected prompts without subsequent task rebalancing.

\begin{table}[htbp]
    \centering
    \small
    \caption{Prompt selection by model-based difficulty filtering. The three correctness bands partition evaluated prompts; unsampled prompts are reported separately.}
    \label{tab:rl_filter_funnel}
    \begin{tabular}{lr}
        \toprule
        Filtering stage & Prompts \\
        \midrule
        Audited candidates & 46{,}250 \\
        Evaluated with eight responses & 35{,}609 \\
        Excluded: zero correct & 11{,}126 \\
        Excluded: seven or eight correct & 12{,}312 \\
        Retained: one to six correct & 12{,}171 \\
        Unsampled at cutoff & 10{,}641 \\
        \bottomrule
    \end{tabular}
\end{table}

\paragraph{Checking use of omics inputs.}
For PRISM primary and secondary activity, we additionally compare correct-profile accuracy with two controls: a matched different-cell profile and no expression profile, retaining the same question, drug, and target. Each task is probed on up to 512 deterministically selected questions. Admission requires at least 128 complete paired probes and an accuracy advantage of at least 0.02 over each control, with a positive lower endpoint of a 95\% paired-bootstrap interval (10{,}000 resamples). Both PRISM tasks pass, so this gate removes no additional selected prompts. The primary-task advantage is 0.044 (interval 0.019--0.070) over the different-cell control and 0.168 (0.143--0.195) over the no-profile control, with 389 pairs; the secondary-task advantages are 0.029 (0.006--0.053) and 0.193 (0.169--0.218), with 408 pairs. These are input-dependence diagnostics for these tasks, not evidence of causal biological reasoning across the entire mixture.

\paragraph{Reward composition.}
Rewards are computed by deterministic task-specific functions; no learned preference reward model or language-model judge is used. For the shared reward function used by the biological tasks, GSM8K, multiple-choice questions, and instruction constraints, let $O\in[0,1]$ be the outcome score, $P\in[0,1]$ the evidence-grounding score, $F\in\{0,1\}$ the valid-format indicator, and $L\in[0,1]$ the reasoning-length penalty. After terminal rejection gates, the reward is
\begin{align}
    \widetilde R &= \operatorname{clip}_{[0,1]}
       \bigl(w_O O + w_P P + 0.05F - 0.05L\bigr), \\
    R &= \begin{cases}
        \widetilde R, & c=1,\\
        \min(\widetilde R,0.25), & c=0.
    \end{cases}
    \label{eq:rl_reward}
\end{align}
The default weights are $(w_O,w_P)=(0.75,0.20)$. Marker-label examples without evidence metadata use $(0.95,0)$. For the remaining tasks, an empty evidence ledger gives $P=0$ while retaining the default weights, so the reward before the length penalty is bounded above by 0.80. These task-dependent reward scales are preserved when computing the unnormalized group advantages below.

For a rationale with $n$ whitespace-delimited words and task budget $B$, $L=\min(\max((n-B)/B,0),1)$. Budgets are 400 words for graph paths; 300 for gene sets and GSM8K; 200 for GDSC, essentiality, PerturbQA, PRISM, and instruction constraints; 180 for DrugMechDB; 160 for biological labels and markers; and 120 for general multiple choice. A long rationale can lose at most 0.05 reward. When both outcome and evidence scores are zero, a well-formed answer earns at most the 0.05 format bonus; incorrect answers with partial credit remain capped at 0.25.

The evidence term checks citations against a per-example ledger. Each claim clause must cite known evidence identifiers, and a lexical checker must find agreement with the corresponding stored values; unknown citations, unsupported claims, or uncited claim clauses set $P=0$. Otherwise, $P=\min(k/2,1)$, where $k$ is the number of distinct supported facts. Nonempty ledgers occur in GDSC, essentiality, and PerturbQA records. This is a limited rule-based grounding check, not a semantic proof that a reasoning trace is faithful. The shared reward function rejects rationales containing a fixed list of leakage-like strings: \texttt{ic50}, \texttt{pic50}, \texttt{zscore}, \texttt{z-score}, \texttt{auc\_published}, ``ground truth,'' ``ground-truth,'' ``the correct answer,'' ``the label is,'' ``prevalence,'' ``is essential in,'' and ``\% of cell lines.'' This conservative lexical gate can also reject legitimate wording; it is not a general leakage detector.

\paragraph{Task-specific outcome verifiers.}
Drug-sensitivity, perturbation, PRISM, mechanism-of-action, and general multiple-choice tasks use exact label or entity matching after normalization. GDSC and essentiality rewards add partial credit from the stored potency margin or continuous dependency target to correct labels. Biological label rewards are weighted by a capped inverse class prior. Gene lists are scored by intersection-over-union, and graph paths by validity against known edges, endpoints, knockout constraints, and shortest-path length, with capped partial credit. GSM8K compares the final number, and instruction-following rewards check one of 11 executable output constraints. The two UltraData families use a separate binary reward with symbolic or alias-based answer matching.

\paragraph{GRPO objective and optimization.}
We use critic-free GRPO~\citep{shao2024deepseekmath} in VERL~\citep{sheng2024hybridflow}. At each update, the rollout policy generates eight responses for each of 32 prompts. The sequence-level reward is attached to the final response token, and the group-centered advantage is broadcast across generated tokens:
\begin{equation}
    A_i=R_i-\frac{1}{G}\sum_{j=1}^{G}R_j,\qquad G=8.
    \label{eq:rl_advantage}
\end{equation}
Unlike the standard deviation-normalized variant of GRPO, our configuration does not divide by the within-group reward standard deviation. For token probability ratio $\rho_{i,t}=\pi_\theta(y_{i,t}\mid x,y_{i,<t})/\pi_{\rm old}(y_{i,t}\mid x,y_{i,<t})$, define
\begin{align}
    s(\rho,A)&=\min\{\rho A,\operatorname{clip}(\rho,0.8,1.2)A\},\\
    s_{\rm dc}(\rho,A)&=\begin{cases}
        \max\{s(\rho,A),3A\}, & A<0,\\
        s(\rho,A), & A\geq0.
    \end{cases}
\end{align}
We minimize the token-averaged loss
\begin{equation}
    \mathcal L=-\langle s_{\rm dc}(\rho,A)\rangle
      +0.01\langle\widehat D_{\rm KL}(\pi_\theta,\pi_{\rm SFT})\rangle
      -0.002\langle H(\pi_\theta)\rangle,
    \label{eq:rl_objective}
\end{equation}
using VERL's low-variance sampled KL estimator and the frozen SFT initializer as reference. KL is an explicit actor-loss term, not part of the task reward or filtering correctness. The entropy term encourages continued exploration. No value network is trained. Table~\ref{tab:rl_hyperparameters_9b} provides the remaining settings.

\begin{table}[htbp]
    \centering
    \small
    \caption{9B GRPO settings, initialized from the model after two SFT epochs. The optimization budget is distinguished from the realized training duration below.}
    \label{tab:rl_hyperparameters_9b}
    \begin{tabular}{ll}
        \toprule
        Setting & Value \\
        \midrule
        Hardware & One node, eight H100 80GB GPUs \\
        Parallelism & Eight-way FSDP; rollout tensor parallelism 1 \\
        Precision / memory & BF16 compute; FP32 master weights; CPU offload \\
        Memory controls & Parameter/optimizer offload; gradient checkpointing \\
        Prompts / responses per update & 32 / 256 (eight responses per prompt) \\
        PPO minibatch / epochs per rollout & 32 prompts / 1 \\
        Dynamic microbatch budget & 4{,}096 tokens per GPU \\
        Prompt / response limit & 1{,}024 / 2{,}048 tokens \\
        Maximum profiles per prompt & 8 \\
        Sampling temperature / top-$p$ / top-$k$ & 1.2 / 1.0 / disabled \\
        Random seed & 42 \\
        Optimizer & AdamW, $\beta=(0.9,0.999)$; weight decay 0.01 \\
        Learning rate & $10^{-6}$, constant; no warmup \\
        Gradient-norm clipping & 1.0 \\
        PPO ratio clip / dual clip & 0.2 / 3.0 \\
        KL / entropy coefficient & 0.01 / 0.002 \\
        Group standard-deviation normalization & Disabled \\
        Training budget & 2 epochs \\
        Training updates & 380 per epoch; 760 budgeted total \\
        Monitoring & 2{,}591 prompts, greedy decoding, every 25 updates \\
        \bottomrule
    \end{tabular}
\end{table}

\paragraph{Training procedure and evaluation.}
The procedure consists of source-data and split verification, validation of omics inputs and reference answers, difficulty filtering using the SFT policy, and on-policy optimization on the fixed set of retained prompts. The trained policy is evaluated using the benchmark protocol. Implementation checks compare rollout and training-policy probabilities, verify finite gradients, and confirm that saved parameters reproduce text-only and omics-conditioned forward passes.

Each epoch drops the final incomplete prompt batch, giving $\lfloor12{,}171/32\rfloor=380$ updates and a two-epoch budget of 760 updates. Monitoring uses greedy decoding with one response per prompt. Training stopped after 724 updates, before completing the budget; the latest retained policy is the checkpoint at update~700.

\subsection{Ablation Studies}
\label{app:ablations}

\begin{table}[t]
  \centering
  \begin{minipage}{\linewidth}
  \small
  \newcommand{\ablmeansize}{\small}
  \newcommand{\ablstdsize}{\scriptsize}
  \newcommand{\ablhead}[1]{{\scriptsize #1}}
  \newcommand{\ablstd}[2]{\begin{tabular}[c]{@{}r@{}}{\ablmeansize #1}\\[-0.8ex]\scalebox{0.92}{{\ablstdsize $\pm$ #2}}\end{tabular}}
  \newcommand{\ablstdbf}[2]{\begin{tabular}[c]{@{}r@{}}{\ablmeansize\bfseries #1}\\[-0.8ex]\scalebox{0.92}{{\ablstdsize $\pm$ #2}}\end{tabular}}
  \refstepcounter{table}
  \label{tab:ablation_results}
  \textbf{Table~\thetable: Ablation study results.}
  The baseline is Qwen3.5-0.8B trained through Pretraining $\rightarrow$ SFT,
  with normalized expression, Funomics T0 and Geneformer embeddings, the count
  indicator, and gene tokens initialized from their mean subword embeddings
  (Section~\ref{sec:method_arch}). Each block repeats this baseline and varies the
  named component, except the RL subgroup, which compares the 9B checkpoint after
  two SFT epochs with the final \modelname{} model ($\dagger$), reported in
  Tables~\ref{tab:tabula_geo_disease}, \ref{tab:geo_qa}, and~\ref{tab:perturbqa_direction}. Scores are mean $\pm$
  standard deviation over three independent training runs; the 9B results are from
  a single run.
  Cell2Sentence-like replaces continuous omics inputs with text listing the
  1{,}024 most highly expressed genes, ranked by expression.
  Scores are scaled to 0--100: macro AUROC for Tabula Sapiens and Human Diseases, mean AUROC
  for PerturbQA, and binary accuracy/free-text judge score for GEO-OmicsQA.
  Overall averages the seven benchmark scores. Higher is better; bold marks the
  best means within each comparison group, including displayed ties.

  \centering
  \resizebox{\linewidth}{!}{%
  \begin{tabular}{lr*{7}{r}}
    \toprule
    & & \multicolumn{2}{c}{Tabula Sapiens} & \multicolumn{2}{c}{Human Diseases} & \multicolumn{2}{c}{GEO-OmicsQA} & \multicolumn{1}{c}{PerturbQA} \\
    \cmidrule(lr){3-4}\cmidrule(lr){5-6}\cmidrule(lr){7-8}\cmidrule(lr){9-9}
    Model & \ablhead{Overall} & \ablhead{Cell type top-20} & \ablhead{Tissue} & \ablhead{Disease subtype} & \ablhead{Tissue} & \ablhead{Binary QA} & \ablhead{Free text} & \ablhead{Mean AUROC} \\
    \midrule
    \multicolumn{9}{l}{\textit{Embedding source}} \\
    Full embedding stack & \ablstdbf{77.6}{0.04} & \ablstd{95.3}{0.03} & \ablstd{75.7}{0.08} & \ablstd{85.4}{0.05} & \ablstdbf{98.0}{0.03} & \ablstd{72.6}{0.17} & \ablstdbf{55.2}{0.15} & \ablstdbf{60.7}{0.13} \\
    No expression embedding & \ablstd{76.2}{0.05} & \ablstdbf{96.7}{0.03} & \ablstd{70.2}{0.07} & \ablstd{83.9}{0.04} & \ablstd{96.3}{0.12} & \ablstdbf{72.9}{0.17} & \ablstd{54.8}{0.20} & \ablstd{58.5}{0.27} \\
    No Funomics T0 embedding & \ablstd{77.2}{0.12} & \ablstd{95.1}{0.04} & \ablstd{75.5}{0.22} & \ablstd{84.2}{0.15} & \ablstdbf{98.0}{0.01} & \ablstd{72.1}{0.47} & \ablstdbf{55.2}{0.20} & \ablstd{60.2}{0.38} \\
    No Geneformer embedding & \ablstd{77.3}{0.14} & \ablstd{92.9}{0.04} & \ablstdbf{77.5}{0.03} & \ablstdbf{86.3}{0.10} & \ablstdbf{98.0}{0.06} & \ablstd{71.2}{0.38} & \ablstd{54.8}{0.26} & \ablstd{60.3}{0.42} \\
    Cell2Sentence-like & \ablstd{71.5}{0.15} & \ablstd{93.6}{0.10} & \ablstd{75.6}{0.15} & \ablstd{77.8}{0.17} & \ablstd{91.2}{0.25} & \ablstd{62.4}{0.42} & \ablstd{51.3}{0.31} & \ablstd{48.8}{0.39} \\
    \midrule
    \multicolumn{9}{l}{\textit{Gene tokenization}} \\
    Gene tokens & \ablstdbf{77.6}{0.04} & \ablstdbf{95.3}{0.03} & \ablstd{75.7}{0.08} & \ablstdbf{85.4}{0.05} & \ablstdbf{98.0}{0.03} & \ablstdbf{72.6}{0.17} & \ablstd{55.2}{0.15} & \ablstdbf{60.7}{0.13} \\
    Random gene initialization & \ablstd{75.6}{0.04} & \ablstdbf{95.3}{0.03} & \ablstd{74.0}{0.15} & \ablstd{83.4}{0.14} & \ablstdbf{98.0}{0.05} & \ablstd{71.9}{0.38} & \ablstd{55.2}{0.11} & \ablstd{51.4}{0.29} \\
    No gene tokens & \ablstd{77.0}{0.01} & \ablstd{95.2}{0.07} & \ablstdbf{78.9}{0.29} & \ablstd{82.8}{0.09} & \ablstd{97.0}{0.03} & \ablstd{70.7}{0.20} & \ablstdbf{56.9}{0.22} & \ablstd{57.7}{0.24} \\
    \midrule
    \multicolumn{9}{l}{\textit{Stages}} \\
    0.8B Pretraining $\rightarrow$ SFT & \ablstdbf{77.6}{0.04} & \ablstdbf{95.3}{0.03} & \ablstd{75.7}{0.08} & \ablstdbf{85.4}{0.05} & \ablstdbf{98.0}{0.03} & \ablstdbf{72.6}{0.17} & \ablstdbf{55.2}{0.15} & \ablstdbf{60.7}{0.13} \\
    0.8B Pretraining only & \ablstd{75.3}{0.20} & \ablstd{95.0}{0.39} & \ablstdbf{80.1}{2.13} & \ablstd{82.2}{1.66} & \ablstd{95.9}{0.67} & \ablstd{66.7}{1.85} & \ablstd{55.1}{0.36} & \ablstd{52.2}{0.97} \\
    0.8B SFT only & \ablstd{73.4}{0.28} & \ablstd{95.2}{0.40} & \ablstd{74.3}{1.61} & \ablstd{81.5}{1.91} & \ablstd{96.1}{0.68} & \ablstd{67.2}{0.76} & \ablstd{49.4}{0.49} & \ablstd{49.8}{0.82} \\
    \noalign{\vskip 2pt}\multispan{9}\leaders\hbox{\rule{2pt}{0.3pt}\hspace{2pt}}\hfill\cr\noalign{\vskip 2pt}
    9B Pretraining $\rightarrow$ SFT & 81.9 & 92.5 & 75.6 & \textbf{90.7} & 98.2 & 75.5 & \textbf{65.1} & 75.7 \\
    9B Pretraining $\rightarrow$ SFT $\rightarrow$ RL$^{\dagger}$ & \textbf{82.4} & \textbf{92.6} & \textbf{76.7} & 90.6 & \textbf{98.4} & \textbf{76.7} & \textbf{65.1} & \textbf{76.4} \\
    \midrule
    \multicolumn{9}{l}{\textit{Backbone size}} \\
    0.8B & \ablstd{77.6}{0.04} & \ablstdbf{95.3}{0.03} & \ablstd{75.7}{0.08} & \ablstd{85.4}{0.05} & \ablstd{98.0}{0.03} & \ablstd{72.6}{0.17} & \ablstd{55.2}{0.15} & \ablstd{60.7}{0.13} \\
    2B & \ablstd{79.9}{0.02} & \ablstd{94.1}{0.08} & \ablstdbf{76.0}{0.05} & \ablstd{85.5}{0.16} & \ablstd{98.1}{0.04} & \ablstd{74.9}{0.17} & \ablstd{60.0}{0.29} & \ablstd{70.9}{0.29} \\
    9B & \textbf{81.9} & 92.5 & 75.6 & \textbf{90.7} & \textbf{98.2} & \textbf{75.5} & \textbf{65.1} & \textbf{75.7} \\
    \bottomrule
  \end{tabular}}
  \end{minipage}
\end{table}

The ablations support our choice of the full embedding stack, pretraining
followed by SFT, and gene tokens initialized from subword embeddings, each of
which achieves the best aggregate score within its comparison group. Larger
backbones generally perform better, although gains are not uniform across tasks. The final 9B checkpoint improves
the overall score from 81.9 to 82.4 after RL, with gains on five of seven metrics
and essentially unchanged performance on the remaining two.

\subsection{GEO-OmicsQA Examples}
\label{app:geo_omicsqa_examples}

One representative question--answer pair for each GEO-OmicsQA benchmark variant is shown below.
The placeholder \texttt{<omics>} denotes a transcriptomic expression embedding injected at that position; $\times N$ indicates $N$ replicate sample embeddings.

\newcommand{\qapair}[2]{%
  \begin{examplebox}
    \textbf{Question:}~#1\par\smallskip
    \textbf{Answer}:~#2
  \end{examplebox}
}

\subsubsection*{Binary QA (standard)}

\qapair{%
  Does the addition of PIK3CA*H1047R (samples \texttt{<omics>}$\times 3$) further decrease E-cadherin (CDH1) expression compared to ARID1A knockdown alone (samples \texttt{<omics>}$\times 3$)?%
}{%
  No.%
}

\subsubsection*{Binary QA (minimal)}

\qapair{%
  Are Ets-family genes like ETV1, ETV4, and ETV5 unaffected in \texttt{<omics>}$\times 2$ compared to \texttt{<omics>}$\times 2$?%
}{%
  No.%
}

\subsubsection*{Simple free-text (standard)}

\qapair{%
  I am analyzing the transcriptomes of non-hemogenic endothelial cells (\texttt{<omics>}$\times 2$) and hemogenic endothelial cells (\texttt{<omics>}$\times 2$) isolated from the AGM region of 30- and 32-day-old human embryos based on ACE expression. What does the expression data indicate about the arterial versus venous identity of the hemogenic subset compared to the non-hemogenic subset?%
}{%
  Both subsets express pan-endothelial markers such as CD34, CDH5, PECAM1, and TEK. However, the hemogenic samples exhibit a clear upregulation of genes classically associated with an arterial fate, including BMX, GJA5, DLL4, CXCR4, and HEY2. Conversely, these hemogenic samples display a lower expression of venous-associated genes like NR2F2 and FLRT2 relative to the non-hemogenic controls. This suggests that the hemogenic subset possesses a distinct arterial gene signature.%
}

\subsubsection*{Simple free-text (minimal)}

\qapair{%
  Examine the EMT process in \texttt{<omics>}$\times 2$ versus \texttt{<omics>}$\times 2$. What do the epithelial and mesenchymal markers reveal?%
}{%
  The first group displays a clear upregulation of the master EMT transcription factor SNAI1, accompanied by a strong downregulation of the epithelial marker E-cadherin (CDH1) compared to the second group. Furthermore, there is a modest increase in N-cadherin (CDH2) expression. This coordinated shift from an epithelial to a mesenchymal transcriptional signature strongly supports the conclusion that the perturbation acts as a triggering signal for the initiation of the transition.%
}

\subsubsection*{Contextual free-text (standard)}

\qapair{%
  During perforated-patch current-clamp recordings, we observed that the Venus+ cells display spontaneous action potential firing. What specific voltage-gated channels expressed in the Venus+ dataset (\texttt{<omics>}$\times 4$) compared to the Venus- dataset (\texttt{<omics>}$\times 4$) provide the molecular basis for this excitability?%
}{%
  The intrinsic excitability you observed is supported by a robust profile of voltage-gated ion channels in the transcriptomic data. The Venus-positive cells express the voltage-gated sodium channel SCN3A, which is vital for the rapid depolarizing phase of action potentials. Furthermore, they are enriched for specific voltage-gated calcium channels, particularly the P/Q-type CACNA1A and the L-type CACNA1C, which contribute to the action potential duration and trigger intracellular calcium-dependent signaling. Alongside the voltage-gated potassium channel KCNB2 for repolarization, this coordinated suite of ion channels forms the molecular foundation for the spontaneous electrical activity you recorded.%
}

\subsubsection*{Contextual free-text (minimal)}

\qapair{%
  We noted a significant enrichment of oxidative phosphorylation genes (p = 0.02) in our resistant model. Using the RNA-seq data for \texttt{<omics>}$\times 3$ compared to \texttt{<omics>}$\times 3$, how is this enrichment reflected at the level of individual respiratory chain complex genes?%
}{%
  The enrichment is distinctly visible across multiple crucial components of the mitochondrial respiratory chain in the first group of samples compared to the second. Specifically, genes encoding subunits of Complex I (NADH dehydrogenase), such as NDUFA1 and NDUFB2, show significant upregulation. Furthermore, multiple subunits associated with Complex IV (cytochrome c oxidase), including COX6C, COX7A2, COX7C, and the assembly factor COX17, are expressed at elevated levels. This coordinated upregulation across different complexes suggests a comprehensive enhancement of the oxidative phosphorylation machinery.%
}

\subsection{Task Examples}
\label{app:task_examples}
Table~\ref{tab:task_examples} shows representative examples from each category of the 157 instruction task groups.
Long questions and answers are truncated.
The placeholder \texttt{<omics>} denotes an omics expression embedding injected at that position.

\definecolor{rowgray}{gray}{0.94}
\definecolor{cathdr}{gray}{0.80}
\newcolumntype{Q}{>{\RaggedRight\arraybackslash\scriptsize}p{0.40\linewidth}}
\newcolumntype{A}{>{\RaggedRight\arraybackslash\scriptsize}p{0.25\linewidth}}
\newcolumntype{T}{>{\RaggedRight\arraybackslash\scriptsize\bfseries}p{0.15\linewidth}}
\newcolumntype{Y}{>{\RaggedRight\arraybackslash\scriptsize}p{0.10\linewidth}}
\newcolumntype{M}{>{\RaggedRight\arraybackslash\scriptsize}p{\dimexpr0.65\linewidth+2\tabcolsep\relax}}
{\setlength{\tabcolsep}{4pt}
\begin{longtable}{T Y Q A}
  \caption{Representative training examples by task category. The Stage 2 instruction corpus is organised into 157 task groups by biological objective and input/output format, in 13 categories; each category header gives its number of groups and conversation records (5,210,786 in total). We show a few representative groups per category, each with its number of training samples and one question--answer example. Two final rows illustrate free-text and text-only sources. Examples are excerpts: long text and lists may be abbreviated, empty thinking tags are omitted, and <omics> denotes an expression input.}
  \label{tab:task_examples} \\
  \toprule
  \normalfont\textbf{Task} & \normalfont\textbf{Type} & \normalfont\textbf{Question} & \normalfont\textbf{Answer} \\
  \midrule
  \endfirsthead
  \multicolumn{4}{l}{\scriptsize\textit{(continued from previous page)}} \\[2pt]
  \toprule
  \normalfont\textbf{Task} & \normalfont\textbf{Type} & \normalfont\textbf{Question} & \normalfont\textbf{Answer} \\
  \midrule
  \endhead
  \midrule
  \multicolumn{4}{r}{\scriptsize\textit{(continued on next page)}} \\
  \endfoot
  \bottomrule
  \endlastfoot
  \rowcolor{cathdr}\multicolumn{4}{l}{\scriptsize\textbf{Single-cell annotation (13 groups; 1,051,316 samples)}} \\
  \addlinespace[1pt]
  Cell-cycle phase\newline{\normalfont 99,969 samples} & scRNA-seq & <omics> Based on the measured expression data, identify the cell cycle phase of the cell. Possible answers: 'S', 'G2M', 'G1' & G1 \\
  \rowcolor{rowgray}Cell-type annotation (with optional context)\newline{\normalfont 479,838 samples} & scRNA-seq & Identify the cell type of <omics> based on the transcriptional profile. & perivascular cell \\
  Malignant-cell identification\newline{\normalfont 8,000 samples} & Tumour scRNA-seq & Use the transcriptional profile to decide whether <omics> is a cancer cell. Answer with "Yes" or "No". & No \\
  \addlinespace[1pt]
  \rowcolor{cathdr}\multicolumn{4}{l}{\scriptsize\textbf{Bulk expression annotation (12 groups; 391,574 samples)}} \\
  \addlinespace[1pt]
  Cell-line identification\newline{\normalfont 52,191 samples} & Bulk RNA-seq & Which cell line does the expression profile support for <omics>? & MSN08 \\
  \rowcolor{rowgray}Perturbation-target identification\newline{\normalfont 69,007 samples} & Bulk RNA-seq & Which target was perturbed for <omics> according to the expression data? & Dexamethasone \\
  GTEx tissue (with optional age or sex)\newline{\normalfont 46,977 samples} & Bulk RNA-seq & Given the expression profile, determine the anatomical tissue for <omics>. Possible answers: "Heart - Left Ventricle", "Heart - Atrial Appendage" & Heart - Left Ventricle \\
  \addlinespace[1pt]
  \rowcolor{cathdr}\multicolumn{4}{l}{\scriptsize\textbf{Cancer and clinical annotation (25 groups; 189,757 samples)}} \\
  \addlinespace[1pt]
  Cancer diagnosis (with optional context)\newline{\normalfont 32,500 samples} & Bulk RNA-seq & This sample comes from Ovary. Given the expression data, identify the disease category of <omics>. & Cystic, Mucinous and Serous Neoplasms \\
  \rowcolor{rowgray}Overall tumour stage (with optional context)\newline{\normalfont 11,729 samples} & Bulk RNA-seq & Based on <omics>, what is the pathological stage of the sample? & Stage IV \\
  Patient specimen versus model\newline{\normalfont 826 samples} & Cancer-model RNA-seq & Does this sample come from a patient tumour rather than a laboratory model, based on the measured expression data? <omics> Answer with 'No' or 'Yes'. & No \\
  \addlinespace[1pt]
  \rowcolor{cathdr}\multicolumn{4}{l}{\scriptsize\textbf{Expression interpretation and comparison (7 groups; 362,886 samples)}} \\
  \addlinespace[1pt]
  GEO binary biological QA\newline{\normalfont 78,828 samples} & Bulk RNA-seq & Review the data for <omics> and <omics>. Does the factor depleted in <omics> normally act to prevent the extension of upstream antisense transcripts? Reply with 'Yes' or 'No'. & Yes \\
  \rowcolor{rowgray}GEO open-ended biological QA\newline{\normalfont 23,591 samples} & Bulk RNA-seq & I have identified a signature of 42 specific transcripts that differentiate these pre-perfusion allografts destined for delayed graft function (<omics>, <omics>, <omics>) from those with immediate function, after adjusting for reperfusion effects. Based on the expression of these 42 genes in the provided data, what upstream biological processes appear to be driving this impaired physiological resilience? & Evaluating the expression of these 42 signature transcripts reveals a strong enrichment for innate immune system activation and cell-mediated immune responses. The upregulation of transcripts such as UBD, NLRC4, and interferon-gamma inducible targets like CXCL10 and [\ldots{}] \\
  Nearest expression-profile retrieval\newline{\normalfont 45,021 samples} & Expression & Given cell lines LN18<omics>, HKA1<omics>, SBC5<omics>, LS513<omics>, KYSE270<omics>, which of them is the most similar to HARA<omics> in terms of expression profile? & KYSE270 \\
  \addlinespace[1pt]
  \rowcolor{cathdr}\multicolumn{4}{l}{\scriptsize\textbf{Chemical and genetic perturbation (13 groups; 530,681 samples)}} \\
  \addlinespace[1pt]
  Downregulated genes after treatment\newline{\normalfont 9,775 samples} & Chemical perturbation & If exposure to a compound targeting EGFR (SMILES COc1cc2ncnc(Nc3c\allowbreak{}cc(F)\allowbreak{}c(Cl)\allowbreak{}c3)\allowbreak{}c2c\allowbreak{}c1OCCCN1CCOCC1)\allowbreak{} at 0.05 $\mu$M is applied in <omics> , what would be the top 59 most downregulated genes? & FN1, ABI3BP, PLXNA2, ADAMTS6, KRT81, SUCNR1, \ldots{} (59 genes in total) \\
  \rowcolor{rowgray}Knocked-out gene identification\newline{\normalfont 46,447 samples} & Genetic perturbation & Given the control <omics> <omics> <omics> <omics> <omics> and the perturbed <omics> <omics> <omics> <omics> <omics> , determine the gene whose knockout caused this change. & SLC25A3 \\
  Perturbation-res\allowbreak{}ponse direction with gene notes\newline{\normalfont 300,000 samples} & Genetic perturbation & Notes: AAR2 = AAR2 splicing factor; 79 partners incl PRPF8, SNRNP200. RPL7A = ribosomal protein L7a; 567 partners. 1 shared partners. Does RPL7A go up or down when AAR2 is knocked down given <omics>? Answer with Increase or Decrease only. & Decrease \\
  \addlinespace[1pt]
  \rowcolor{cathdr}\multicolumn{4}{l}{\scriptsize\textbf{Cancer genomics (5 groups; 21,297 samples)}} \\
  \addlinespace[1pt]
  Copy-number-altered genes\newline{\normalfont 14,824 samples} & Cell-line genomics & Which cancer genes have a copy number gain in cell line WM35, sorted alphabetically? & AFDN AXIN2 BACH2 CD79B DDX5 DEK \ldots{} (16 genes in total) \\
  \rowcolor{rowgray}Mutated cancer genes\newline{\normalfont 3,448 samples} & Cell-line genomics & Which cancer genes are mutated in cell line WM88<omics>, sorted alphabetically? & BRAF: homozygous \\
  \addlinespace[1pt]
  \rowcolor{cathdr}\multicolumn{4}{l}{\scriptsize\textbf{Pathways, gene sets, and molecular knowledge (19 groups; 653,481 samples)}} \\
  \addlinespace[1pt]
  Directed-edge verification\newline{\normalfont 51,157 samples} & Graph / gene set & Is there an 'indirect effect' direct relationship from PRKAG2 to SREBF1? & Reasoning: PRKAG2 has 33 forward neighbour(s): ACACA, ACACB, AKT1S1, CD36, CFTR, CRTC2, CRY1, EEF2K, ELAVL1, FOXO1, FOXO3, FOXO4, \ldots{}. SREBF1 is among them. Final: Yes \\
  \rowcolor{rowgray}Shortest path after knockout\newline{\normalfont 199,758 samples} & Graph / gene set & If the following nodes are knocked out: [[ANK1, CDC42, CEL, PRKCA, PRKCB, TWIST2, VAV3]], what is the shortest path from MIR520E to LIPA in the biological pathways? & Reasoning: Blocked nodes: ANK1, CDC42, CEL, PRKCA, PRKCB, TWIST2, VAV3. Target is LIPA; from MIR520E the shortest route is 9 step(s). From MIR520E: 1 open [\ldots{}] Final: MIR520E inhibit CD44 bind to IQGAP1 activate MAPK1 activate PRKCG indirect effect PRKACA express HSD17B7 sequential reactions DHCR24 produce DHCR7 produce LIPA \\
  Protein function\newline{\normalfont 20,000 samples} & Protein knowledge & What can be a function of the protein MYOD1? & Acts as a transcriptional activator that promotes transcription of muscle-specific target genes and plays a role in muscle differentiation \\
  \addlinespace[1pt]
  \rowcolor{cathdr}\multicolumn{4}{l}{\scriptsize\textbf{Cell and disease ontologies (9 groups; 126,858 samples)}} \\
  \addlinespace[1pt]
  Cell type from definition\newline{\normalfont 2,899 samples} & Ontology & Which cell type is described by the following definition? "A langerin-positive lymph node dendritic cell that is CD103-positive and CD11b-low." & CD103-positive, langerin-positive lymph node dendritic cell \\
  \rowcolor{rowgray}Cell/disease ancestry verification\newline{\normalfont 62,824 samples} & Ontology & Is "superior cervical ganglion TH neuron" a kind of "neural cell"? Answer Yes or No. & Yes \\
  \addlinespace[1pt]
  \rowcolor{cathdr}\multicolumn{4}{l}{\scriptsize\textbf{Expression annotation with reasoning (9 groups; 504,429 samples)}} \\
  \addlinespace[1pt]
  Bulk tissue with reasoning\newline{\normalfont 22,219 samples} & Bulk RNA-seq & I need a quick label for this bulk sample, so what tissue origin does the overall programme in <omics> best support, and why? Add your concise prediction after "Answer:" on a new line. & Reasoning: The profile is dominated by the surfactant metabolism programme, with high detection of SFTPC, SFTPA1, SFTPA2, SFTPB, SFTPD, NAPSA, and SLC34A2. [\ldots{}] Final: Answer: Lung \\
  \rowcolor{rowgray}Single-cell type with reasoning (optional context)\newline{\normalfont 253,738 samples} & scRNA-seq & Which cell type is best supported by the overall expression programme in <omics>? Explain your reasoning using the key markers, and put your prediction after the `Answer:` marker on a separate final line. & Reasoning: The profile is defined by the co-detection of CD3D and CD3E, the structural subunits of the T-cell receptor complex that are essential for [\ldots{}] Final: Answer: t cell \\
  \addlinespace[1pt]
  \rowcolor{cathdr}\multicolumn{4}{l}{\scriptsize\textbf{Essentiality, drug response, and mechanisms (14 groups; 486,695 samples)}} \\
  \addlinespace[1pt]
  Gene essentiality and dependency score\newline{\normalfont 300,000 samples} & Expression / drug & <omics> This RNA expression profile is from a Synovial Sarcoma cell line (Soft Tissue). Cancer lineage: Soft Tissue > Synovial Sarcoma. Target gene AGA LoF: no. Tissue-distinctive LoF genes: AOX1, ARMCX4, CKMT2, COMMD10, DPY19L2, KPNA1, LARS2, MPDZ, NSD2, OR5M1, PDCD6IP, PDS5A, RGS6, RPGRIP1, SH3RF1, STAG2, SZT2, [\ldots{}] least 0.5 and No otherwise. Truncate the value to one decimal without rounding. Return only a JSON object with "essential\_dep" first and "dependancy" second. & \{"essential\_dep": "No", "dependancy": 0.0\} \\
  \rowcolor{rowgray}GDSC drug sensitivity\newline{\normalfont 48,248 samples} & Expression / drug & <omics> This RNA expression profile is from the A101D cell line (Skin, Melanoma). In the GDSC screen, the compound is CAMPTOTHECIN. Its annotated targets are TOP1. Predict the source's final sensitivity label, assigned by its compound-specific GMM/Z-score procedure and corrections. Is this cell line sensitive or resistant to CAMPTOTHECIN? Answer with sensitive or resistant only. & sensitive \\
  Effect of blocking a mechanism step\newline{\normalfont 9,238 samples} & Drug mechanism & Indication: Potassium hydrogencarbonate for Hypokalemia. Suppose this step does not take place. Potassium falls as Hypokalemia rises. What does that do to the drug? No Effect: that is not part of how the drug works, so its mechanism is untouched. Partially Blocked: it is part of the mechanism, but another route to the disease remains. Fully Blocked: it is part of the mechanism and every route runs through it. Answer with the option name only. & Fully Blocked \\
  \addlinespace[1pt]
  \rowcolor{cathdr}\multicolumn{4}{l}{\scriptsize\textbf{Human genetics and disease associations (11 groups; 300,000 samples)}} \\
  \addlinespace[1pt]
  Gene–disease association verification\newline{\normalfont 34,976 samples} & Genetics & Use only the quality-filtered GWAS, rare-variant burden, ClinVar, Gene2Phenotype, Genomics England PanelApp, Orphanet, and ClinGen records retained in this benchmark. In Open Targets 26.06, is there recorded human genetic evidence linking AMY2B to dentures? Answer Yes or No. & Yes \\
  \rowcolor{rowgray}GWAS effector-gene prioritisation\newline{\normalfont 34,976 samples} & Genetics & For the GWAS of Platelet count at 6\_36381023\_C\_T, select the most likely effector gene among these candidates [\ldots{}]. Candidates: [\{"features": \{[\ldots{}] "distanceFootpri\allowbreak{}ntMean":\allowbreak{} 1.0, [\ldots{}]\}, "gene": "ETV7"\}, \{"features": \{[\ldots{}] "distanceFootpri\allowbreak{}ntMean":\allowbreak{} 0.98992, [\ldots{}]\}, "gene": "KCTD20"\}]. Answer with one candidate gene symbol. & ETV7 \\
  \addlinespace[1pt]
  \rowcolor{cathdr}\multicolumn{4}{l}{\scriptsize\textbf{Biological and general instruction following (8 groups; 86,298 samples)}} \\
  \addlinespace[1pt]
  Constrained expression annotation\newline{\normalfont 13,437 samples} & Omics instruction & Given this expression profile, determine the anatomical tissue for it. <omics> Possible answers: "Breast - Mammary Tissue", "Bladder", "Artery - Tibial", [\ldots{}] & Bladder \\
  \rowcolor{rowgray}Constrained biology responses\newline{\normalfont 19,503 samples} & Biology QA & How does the hypothalamus regulate the secretion of hormones from the anterior pituitary gland? Answer in at most 33 words. & The hypothalamus regulates the secretion of hormones from the anterior pituitary gland through a complex system of hormonal and neural connections. This process involves the following steps: 1. \\
  \addlinespace[1pt]
  \rowcolor{cathdr}\multicolumn{4}{l}{\scriptsize\textbf{General capabilities (12 groups; 505,514 samples)}} \\
  \addlinespace[1pt]
  Mathematical reasoning\newline{\normalfont 97,717 samples} & Mathematics & Natalia sold clips to 48 of her friends in April, and then she sold half as many clips in May. How many clips did Natalia sell altogether in April and May? & Reasoning: Natalia sold 48/2 = 24 clips in May. Natalia sold 48+24 = 72 clips altogether in April and May. Final: 72 \\
  \rowcolor{rowgray}General instruction and dialogue\newline{\normalfont 66,734 samples} & Dialogue & How do you know if an argument you come across makes sense logically? Can you walk me through the process you would use to break down an argument and understand it? & To determine whether an argument makes sense logically, I follow a process of breaking it down into its core components, analyzing the relationships between them, and evaluating the evidence provided. Here's how I would walk through this process: When I encounter an [\ldots{}] \\
  Tool invocation and dialogue\newline{\normalfont 29,066 samples} & Tool use & Hi, can you tell me the current price of Apple's stock? & <tool\_call> \{"name": "get\_stock\_price", "arguments": \{"stock\_symbol": "AAPL"\}\} </tool\_call> \\
  \addlinespace[1pt]
  \rowcolor{cathdr}\multicolumn{4}{l}{\scriptsize\textbf{Free-text and text-only sources (examples)}} \\
  \addlinespace[1pt]
  PubMed abstract & Free text & \multicolumn{2}{M}{Investigation of the carbohydrate metabolism of normal and neoplastic hepatocytes using 2,\allowbreak{}6-dichlorophenolin\allowbreak{}dophenol as a probe for NAD(P)H production measured by voltammetry. An electrochemical technique is described for measurement of intracellular NAD(P)H production. This technique involves an auxiliary redox system taken up by the cells which is then measured voltammetrically after reduction by NAD(P)H. The redox syst\ldots{}} \\
  \rowcolor{rowgray}Document relations (BioRED) & Relation extraction & Extract the normalized biomedical entities and document-level relations from this PubMed title and abstract. Group unique surface forms that share a database identifier into one entity. Use null as the id only when the text has no normalized identifier. Relations are undirec\ldots{} & \{"entities":\allowbreak{}[\{"id":\allowbreak{}"D006965",\allowbreak{}"type":\allowbreak{}"DiseaseOrPhenoty\allowbreak{}picFeature",\allowbreak{}"mentions":\allowbreak{}["parathyroid hyperplasia"]\allowbreak{}\}\allowbreak{},\allowbreak{}\{"id":\allowbreak{}"D009369",\allowbreak{}"type":\allowbreak{}"DiseaseOrPhenoty\allowbreak{}picFe\ldots{} \\
  \addlinespace[2pt]
\end{longtable}}

\subsection{Asset Provenance, Licenses, and Terms of Use}
\label{app:licenses}

Table~\ref{tab:licenses} summarizes the licenses or terms reported by upstream
providers and the use of each third-party asset in this work.

\paragraph{GTEx acknowledgement.}
The GTEx data used in this work were obtained from GTEx Analysis Release V10
(dbGaP Accession phs000424.v10.p2).

\paragraph{DepMap provenance.}
DepMap 24Q4 expression and dependency data provide cell-line profiles,
gene-essentiality labels, and genomic annotations. The release's stated license
and the DepMap portal terms are recorded separately from the licenses of other
datasets accessed through the portal, such as PRISM.
We do not redistribute the raw DepMap files.

\definecolor{licensegray}{gray}{0.94}
\definecolor{licensehdr}{gray}{0.80}
\newcolumntype{B}{>{\RaggedRight\arraybackslash\scriptsize\bfseries}p{0.30\linewidth}}
\newcolumntype{C}{>{\RaggedRight\arraybackslash\scriptsize}p{0.27\linewidth}}
\newcolumntype{D}{>{\RaggedRight\arraybackslash\scriptsize}p{\dimexpr0.43\linewidth-4\tabcolsep\relax}}

{\setlength{\tabcolsep}{2pt}
\begin{longtable}{@{}B C D@{}}
    \caption{Third-party assets used in \modelname{} and in evaluation.}
    \label{tab:licenses} \\
    \toprule
    \normalfont\textbf{Asset} & \normalfont\textbf{License / Terms} & \normalfont\textbf{Use in this work} \\
    \midrule
    \endfirsthead
    \multicolumn{3}{l}{\scriptsize\textit{Table~\ref{tab:licenses} continued from previous page}} \\[2pt]
    \toprule
    \normalfont\textbf{Asset} & \normalfont\textbf{License / Terms} & \normalfont\textbf{Use in this work} \\
    \midrule
    \endhead
    \midrule
    \multicolumn{3}{r}{\scriptsize\textit{continued on next page}} \\
    \endfoot
    \bottomrule
    \endlastfoot
    \rowcolor{licensehdr}\multicolumn{3}{l}{\scriptsize\textbf{Expression datasets}} \\
    \addlinespace[1pt]
    Human Cell Atlas~\citep{regev2017science, cellxgene2025census} &
    CC~BY~4.0 &
    Primary training source for single-cell annotation tasks. \\

    \rowcolor{licensegray}Tabula Sapiens~\citep{tabula2022sapiens} &
    CC~BY~4.0 &
    Held-out evaluation benchmark (cell-type annotation); not used in training. \\

    GTEx~\citep{gtex2020atlas} &
    GTEx Portal terms of use &
    Tissue-of-origin prediction (training). Acknowledgement given in this appendix. \\

    \rowcolor{licensegray}TCGA~\citep{weinstein2013tcga} &
    NIH Genomic Data Sharing Policy &
    Clinical \& bulk prediction tasks (training). Open access data only; no controlled-access data used. \\

    GEO (via ARCHS4 pipeline)~\citep{clough2024geo, lachmann2018massive} &
    NCBI/GEO source terms; ARCHS4 pipeline: Apache~2.0 &
    GEO-OmicsQA benchmark expression profiles and GEO sample-description training tasks. \\

    \rowcolor{licensegray}Human Diseases~\citep{schaefer2025cellwhisperer} &
    CC~BY~4.0 &
    Held-out disease-classification evaluation; samples overlapping training removed. \\

    Replogle et al.\ 2022~\citep{replogle2022mapping} &
    CC~BY~4.0 (Figshare+ and scPerturb depositions) &
    Perturbation-response training tasks (Perturb-seq). \\

    \rowcolor{licensegray}Nadig et al.\ 2025~\citep{nadig2025transcriptome} &
    GEO source terms &
    Perturbation-response training tasks (Perturb-seq). \\

    DepMap Expression~\citep{depmap2024, pacini2024depmap} &
    CC~BY~4.0; DepMap portal terms &
    Cell-line expression profiles for gene essentiality and genomics tasks. \\

    \rowcolor{licensegray}DepMap Gene Dependency~\citep{depmap2024} &
    CC~BY~4.0; DepMap portal terms &
    CRISPR gene dependency labels for essentiality prediction tasks. \\

    X-Atlas / Orion (HCT116)~\citep{huang2025xatlas} &
    \textbf{CC~BY-NC-SA~4.0} (non-commercial) &
    Genetic-perturbation response training data. Non-commercial research use only. \\

    \addlinespace[2pt]
    \rowcolor{licensehdr}\multicolumn{3}{l}{\scriptsize\textbf{Additional expression and perturbation sources}} \\
    \addlinespace[1pt]
    3CA~\citep{gavish2023hallmarks} &
    Study-specific terms;  GEO source terms; &
    Cancer single-cell characterization and expression tasks. \\

    \rowcolor{licensegray}CancerModels.org~\citep{perova2022pdcm} &
    EMBL-EBI and provider-specific terms; CC0~1.0 &
    Cancer-model characterization; \\

    Tahoe-100M~\citep{zhang2025tahoe} &
    CC0~1.0 &
    Chemical-perturbation response and dose comparisons. \\

    \rowcolor{licensegray}sci-Plex3~\citep{srivatsan2020sciplex} &
    CC~BY~4.0 (scPerturb deposition) &
    Chemical-perturbation tasks from pseudobulk profiles. \\

    sci-Plex GxE~\citep{mcfalinefigueroa2024sciplexgxe} &
    NCBI molecular-data usage policy &
    Genetic--chemical perturbation tasks from pseudobulk profiles. \\

    \rowcolor{licensegray}Jiang et al.~\citep{jiang2025pathway} &
    CC~BY~4.0 (data deposition) &
    Genetic-perturbation response tasks. \\

    Feng et al.~\citep{feng2025crispri} &
    MIT (count-data deposition) &
    Genetic-perturbation response tasks. \\

    \rowcolor{licensegray}Virtual Cell Challenge\newline{\normalfont \url{https://github.com/ArcInstitute/arc-virtual-cell-atlas/tree/main/virtual-cell-challenge}} &
    CC0~1.0 &
    Genetic-perturbation profiles. \\

    CellWhisperer conversations~\citep{schaefer2025cellwhisperer} &
    No explicit license; publicly released by the authors as LLM-curated training data (paper data availability statement) &
    Expression-grounded conversation training; distinct from the held-out Human Diseases dataset. Used for research only; not redistributed. \\

    \rowcolor{licensegray}PerturbQA~\citep{wu2025perturbqa, wu2025perturbqadata} &
    CC~BY~4.0 (published dataset) &
    Perturbation detection and response reasoning; released baseline predictions (\texttt{results.zip}) rescored for evaluation. \\

    \addlinespace[2pt]
    \rowcolor{licensehdr}\multicolumn{3}{l}{\scriptsize\textbf{Screens, drug response, and human genetics}} \\
    \addlinespace[1pt]
    AssayBench~\citep{debrouwer2026assaybench} &
    MIT (dataset release) &
    Ranking and identifying phenotypic-screen hits. \\

    \rowcolor{licensegray}GDSC~\citep{yang2013gdsc} &
    GDSC release 8.5 data-use terms &
    Drug sensitivity and IC50 prediction. \\

    PRISM~\citep{corsello2020prism} &
    CC~BY~4.0 (19Q4 deposition); DepMap portal terms &
    Sensitivity and secondary-screen activity prediction. \\

    \rowcolor{licensegray}DrugMechDB~\citep{gonzalezcavazos2023drugmechdb} &
    CC0~1.0 &
    Mechanism paths and counterfactual reasoning after benchmark exclusions. \\

    Open Targets~\citep{ghoussaini2021otg} &
    Platform data: CC0~1.0; retain source attribution &
    Gene--disease, variant, and GWAS-locus tasks. \\

    \addlinespace[2pt]
    \rowcolor{licensehdr}\multicolumn{3}{l}{\scriptsize\textbf{Biological databases and ontologies}} \\
    \addlinespace[1pt]
    GENCODE~\citep{mudge2025gencode} &
    CC~BY~4.0 &
    Defines the 19,260-gene panel. \\

    \rowcolor{licensegray}Gene Ontology~\citep{ashburner2000go, geneontology2025} &
    CC~BY~4.0 &
    Gene-set annotations for pathway reasoning tasks (training). \\

    Cellosaurus~\citep{bairoch2018cellosaurus} &
    CC~BY~4.0 &
    Cell-line annotations (disease, lineage, transformant) for genomics tasks (training). \\

    \rowcolor{licensegray}UniProt (Swiss-Prot)~\citep{uniprot2025} &
    CC~BY~4.0 &
    Protein function, localization, and keyword annotations for protein QA tasks (training). \\

    BioReason-Pro SFT Reasoning Data~\citep{fallahpour2026bioreasonpro} &
    Apache~2.0 (dataset card) &
    15{,}000 protein-function reasoning conversations for Stage 2. \\

    \rowcolor{licensegray}STRING~\citep{szklarczyk2023string} &
    CC~BY~4.0 &
    Protein--protein interaction graph for network reasoning tasks (training). \\

    Human Protein Atlas~\citep{karlsson2021hpasc,uhlen2015hpa} &
    CC~BY~4.0 &
    Cell, immune and tissue evidence for marker-grounded reasoning data. \\

    \rowcolor{licensegray}TISSUES~\citep{palasca2018tissues} &
    CC~BY~4.0 &
    Tissue--gene evidence for marker-grounded reasoning data. \\

    Reactome~\citep{milacic2024reactome} &
    CC0 (pathway data) &
    Published pathway programmes used in marker evidence cards. \\

    \rowcolor{licensegray}HGNC~\citep{seal2023hgnc} &
    CC0~1.0 &
    Approved symbols, aliases and chromosome annotations for marker-grounded reasoning data. \\

    Cell Ontology\newline{\normalfont \url{https://obofoundry.org/ontology/cl.html}} &
    CC~BY~4.0 &
    Cell-type names, synonyms and hierarchy for label normalization. \\

    \addlinespace[2pt]
    \rowcolor{licensehdr}\multicolumn{3}{l}{\scriptsize\textbf{Pathway and graph-reasoning training sources}} \\
    \addlinespace[1pt]
    BioMaze (KEGG pathway data)~\citep{zhao2025biomaze, zhao2025biomaze_graph} &
    KEGG terms of use &
    KEGG-derived graph for reasoning training. \\

    \rowcolor{licensegray}BioReason KEGG~\citep{fallahpour2025bioreason} &
    Apache~2.0 (dataset card) &
    1{,}159 pathway-context reasoning conversations for Stage 2. \\

    MSigDB~\citep{liberzon2015msigdb} &
    CC~BY~4.0 &
    Gene sets for pathway-activity tasks and drug-response context programmes (training). \\

    \addlinespace[2pt]
    \rowcolor{licensehdr}\multicolumn{3}{l}{\scriptsize\textbf{Biomedical QA and information extraction}} \\
    \addlinespace[1pt]
    BioASQ~\citep{krithara2023bioasq} &
    CC~BY~2.5 &
    Biomedical question answering. \\

    \rowcolor{licensegray}ReasonMed / CoTMed / ResponseMed~\citep{sun2025reasonmed} &
    Apache~2.0 (card covers ReasonMed, CoTMed and ResponseMed) &
    Medical answers and reasoning traces serialized as text. \\

    Medical-R1-Distill-Data~\citep{chen2024huatuogpto1} &
    Apache~2.0 &
    Medical QA and distilled reasoning. \\

    \rowcolor{licensegray}MedReason~\citep{wu2025medreason} &
    Apache~2.0 (export card) &
    Medical reasoning text. \\

    M23K~\citep{huang2025m1, thapa2025disentangling} &
    MIT (dataset release) &
    Medical reasoning text \\

    \rowcolor{licensegray}TheBioCollection~\citep{seo2026thebiocollection} &
    Apache~2.0 (corpus card) &
    Free-text and instruction streams; retain underlying resource provenance. \\

    CAMEL Biology~\citep{li2023camel} &
    CC~BY-NC~4.0 &
    Biology instruction examples. \\

    \rowcolor{licensegray}BioRED~\citep{luo2022biored} &
    No explicit archive license; article-specific terms &
    Entity/relation training and held-out evaluation. Restricted to articles whose licenses permit non-commercial reuse. \\

    BioREx: DrugProt / CID~\citep{lai2023biorex} &
    DrugProt: CC~BY~4.0; CID: public-domain mark; aggregate archive: no explicit license &
    Relation-classification training; other BioREx components excluded. Restricted to articles whose licenses permit non-commercial reuse. \\

    \rowcolor{licensegray}ChemProt~\citep{krallinger2017chemprot} &
    Public Domain Mark~1.0 (BigBio dataset metadata) &
    Chemical--protein relations; official train/dev/test separation retained. \\

    \addlinespace[2pt]
    \rowcolor{licensehdr}\multicolumn{3}{l}{\scriptsize\textbf{Scientific and biomedical text and code}} \\
    \addlinespace[1pt]
    PubMed abstracts\newline{\normalfont \url{https://pubmed.ncbi.nlm.nih.gov/}} &
    NLM source terms and abstract copyrights &
    Biomedical text; no blanket public-domain designation for abstracts. \\

    \rowcolor{licensegray}PubMed Central\newline{\normalfont \url{https://pmc.ncbi.nlm.nih.gov/tools/openftlist/}} &
    Article-specific terms; enriched subset selected using CC~BY / CC0 evidence &
    Biomedical full text. Restricted to articles whose licenses permit non-commercial reuse. \\

    arXiv\newline{\normalfont \url{https://info.arxiv.org/help/license/}} &
    Article-specific licenses &
    Scientific text. Restricted to articles whose licenses permit non-commercial reuse. \\

    \rowcolor{licensegray}bioRxiv / medRxiv\newline{\normalfont \url{https://www.biorxiv.org/}} &
    Article-specific licenses &
    Preprint text with captions and tables; source license metadata retained by the exporter. \\

    AACT / ClinicalTrials.gov\newline{\normalfont \url{https://aact.ctti-clinicaltrials.org/}} &
    AACT and ClinicalTrials.gov terms &
    Clinical-trial records rendered as text. \\

    \rowcolor{licensegray}DailyMed\newline{\normalfont \url{https://dailymed.nlm.nih.gov/dailymed/}} &
    NLM terms and source-label rights &
    Drug-label text. \\

    NIH RePORTER\newline{\normalfont \url{https://reporter.nih.gov/exporter}} &
    NIH website reuse policy &
    Research-project abstracts used for non-commercial research. \\

    \rowcolor{licensegray}ChemPile-Code~\citep{mirza2025chempile} &
    Collection: AGPL-3.0; repository-license-label filter: MIT, Apache~2.0, BSD-2/3-Clause &
    Biomedical Python text; original code, notices, and both license levels retained. \\

    \addlinespace[2pt]
    \rowcolor{licensehdr}\multicolumn{3}{l}{\scriptsize\textbf{General-purpose text and instruction data}} \\
    \addlinespace[1pt]
    Wikipedia via RedPajama~\citep{weber2024redpajama} &
    CC~BY-SA~3.0 (upstream Wikipedia card); GFDL where applicable &
    General text from the RedPajama Wikipedia slice (2023-03-20 dump). \\

    \rowcolor{licensegray}Open PerfectBlend~\citep{labonne2024openperfectblend} &
    Apache~2.0 collection; source attribution retained &
    Conversations, including AutoIF, LMSYS, and UltraChat components. \\

    GSM8K / MetaMathQA~\citep{cobbe2021gsm8k, yu2023metamath} &
    MIT (both dataset cards) &
    Mathematical reasoning. \\

    \rowcolor{licensegray}OpenMathReasoning~\citep{moshkov2025openmathreasoning} &
    CC~BY~4.0 &
    Mathematical reasoning traces. \\

    ARC / MMLU~\citep{clark2018arc, hendrycks2020mmlu} &
    ARC: CC~BY-SA~4.0; MMLU: MIT (dataset cards) &
    General knowledge questions. \\

    \rowcolor{licensegray}Magicoder-Evol-Instruct~\citep{wei2023magicoder} &
    Apache~2.0 &
    Coding instructions. \\

    OpenCodeReasoning~\citep{ahmad2025opencodereasoning} &
    CC~BY~4.0; original question-source terms &
    Competitive-programming reasoning traces. \\

    \rowcolor{licensegray}SmolTalk~\citep{allal2025smollm2} &
    Constraints, rewriting, Magpie and SystemChats: Apache~2.0; Self-OSS-Instruct: ODC-BY~1.0 &
    Instruction following, rewriting, coding and system-prompt adherence. \\

    Hermes / ToolACE\newline{\normalfont \url{https://huggingface.co/datasets/NousResearch/hermes-function-calling-v1}} &
    Apache~2.0 (both dataset cards) &
    Tool-use conversations. \\

    \rowcolor{licensegray}When2Call\newline{\normalfont \url{https://huggingface.co/datasets/nvidia/When2Call}} &
    CC~BY~4.0 &
    Tool-use decisions and responses. \\

    xLAM irrelevance\newline{\normalfont \url{https://huggingface.co/datasets/MadeAgents/xlam-irrelevance-7.5k}} &
    CC~BY~4.0 &
    Recognizing when available tools do not answer a request. \\

    \rowcolor{licensegray}Derived instruction components (Appendix~\ref{app:instruction_derivatives}) &
    CAMEL-derived constraints: CC~BY-NC~4.0; omics variants: parent-source terms &
    Constraints, multi-turn conversations, and reformatted omics instructions. BioReason-Pro is listed separately. \\

    UltraData-SFT-Agent~\citep{openbmb2026ultradataagent} &
    Apache~2.0 (dataset card) &
    Agent trajectories and tool interactions. \\

    \addlinespace[2pt]
    \rowcolor{licensehdr}\multicolumn{3}{l}{\scriptsize\textbf{Foundation models and backbones}} \\
    \addlinespace[1pt]
    Funomics T0~\citep{funomics} &
    Explicit permission from the authors for use in this work &
    512-dimensional embeddings used directly as part of every omics input vector, obtained from the authors on request. \\

    \rowcolor{licensegray}Geneformer-V2-104M~\citep{chen2026geneformer} &
    Apache~2.0 &
    Pretrained single-cell encoder; embeddings used as input features. \\

    Qwen3.5~\citep{qwen2026qwen35} &
    Apache~2.0 &
    Pretrained text backbone; jointly trained with the omics projection. Also evaluated as a GEO-OmicsQA baseline. 0.8B and 2B backbones for ablations. \\

    \addlinespace[2pt]
    \rowcolor{licensegray}Gemini 3.5 Flash-Lite~\citep{google2026gemini35flashlite} &
    Gemini API terms &
    Mechanism-of-action data generation (Appendix~\ref{app:moa_dataset}). \\

    Qwen3.8-27B~\citep{qwen2026qwen38} &
    Apache~2.0 &
    Question and rationale generation for marker-grounded reasoning data. Also evaluated as a GEO-OmicsQA baseline. \\

    \rowcolor{licensegray}Gemma-4-31B-it~\citep{google2026gemma4} &
    Apache~2.0 &
    Quality grading of marker-grounded rationales. \\

    Gemini 3.1 Flash-Lite~\citep{google2026gemini31flashlite} &
    Gemini API terms &
    DeepEval judge model for GEO-OmicsQA free-text scoring. \\

    \addlinespace[2pt]
    \rowcolor{licensehdr}\multicolumn{3}{l}{\scriptsize\textbf{Baseline models}} \\
    \addlinespace[1pt]
    CellWhisperer~\citep{schaefer2025cellwhisperer} &
    CC~BY~4.0 &
    Evaluation baseline: we evaluate the released chat model (LM rows); CLIP rows are taken from the published paper. Profile-linked training conversations are listed separately above. \\

    \rowcolor{licensegray}C2S-Scale-27B~\citep{rizvi2025scaling} &
    CC~BY~4.0 &
    Evaluation baseline (zero-shot annotation, GEO-OmicsQA). \\

    Cell-o1~\citep{fang2025cello1} &
    No standard license (model card: \texttt{other}); NLM disclaimer &
    Evaluation baseline (zero-shot annotation, GEO-OmicsQA). \\

    \rowcolor{licensegray}TxGemma-27B~\citep{wang2025txgemma} &
    Gemma Terms of Use &
    Evaluation baseline (GEO-OmicsQA). \\

    Gemini 3.1 Pro / 3.8 Flash~\citep{google2026gemini31pro,google2026gemini38flash} &
    Google APIs Terms of Service &
    Evaluation baseline (GEO-OmicsQA). \\

    \rowcolor{licensegray}GPT-6 Sol / Luna~\citep{openai2026gpt6soluna} &
    OpenAI Terms of Service &
    Evaluation baseline (GEO-OmicsQA). \\

    AROMA~\citep{wang2026aroma} &
    No explicit license (model card and repository) &
    PerturbQA baseline reproduced from the released checkpoint (evaluation only). \\

    \addlinespace[2pt]
    \rowcolor{licensehdr}\multicolumn{3}{l}{\scriptsize\textbf{Software frameworks}} \\
    \addlinespace[1pt]
    DeepEval\newline{\normalfont \url{https://github.com/confident-ai/deepeval}} &
    Apache~2.0 &
    LLM-as-a-judge framework for GEO-OmicsQA free-text scoring. \\

    \rowcolor{licensegray}\textsc{pdex}~\citep{arcinstitute2025pdex} &
    MIT &
    Parallel differential expression analysis for perturbation-response tasks. \\

    VERL~\citep{sheng2024hybridflow} &
    Apache~2.0 &
    GRPO reinforcement-learning framework (Stage 3). \\

    \rowcolor{licensegray}vLLM~\citep{kwon2023vllm} &
    Apache~2.0 &
    Inference engine for local-model evaluation. \\

\end{longtable}
}

\subsection{Qualitative Examples of Biological Knowledge and Expression Interpretation}
\label{app:capability_examples}

We present nine assessment conversations with \modelname{}. Examples 1--4 illustrate
quantitative biochemical recall, mechanistic explanation, and comparative
interpretation of expression profiles through the same conversational interface.
Examples 5--6 infer a genetic dependency and a cancer-model type from expression
profiles, and Examples 7--9 interpret profiles from real GEO studies.
Each prompt was run once with greedy decoding, no system message, and a fresh conversation history.
The two knowledge questions used text alone; for expression questions,
the chat interface resolved the named cell lines or sample identifiers to their corresponding
20{,}541-feature omics profiles. The prompts below reproduce the user-entered
text, before automatic profile insertion. Answers preserve the generated
wording, with only LaTeX formatting applied.

These are illustrative assessments rather than an aggregate benchmark.
Expression statements were checked against the source data: DepMap 24Q4 values for
the cell-line comparisons (Examples 3--5), the CancerModels.org expression matrix for
Example 6, and our re-processed profiles of the referenced GEO samples (Appendix~\ref{app:datasets}) for
Examples 7--9. In Examples 3--5 the cell-line names were visible to the model; in
Examples 6--9 the sample identifiers were replaced by the profiles, but the question
text still describes the experimental context. Because no anonymized or
swapped-profile controls were run for this checkpoint, these examples do not isolate
the contribution of the omics inputs from prior knowledge of the named cell lines or
of the described experiments.

\subsubsection{Example 1: Quantitative biochemical recall}
\label{app:capability_example_1}

\qapair{%
In the oxidative pentose phosphate pathway, how many NADPH and CO2 molecules are produced per glucose-6-phosphate converted to ribulose-5-phosphate, and is ATP produced directly? Name the two NADPH-producing enzymes and explain why this pathway supports proliferating cancer cells. Keep the answer under 160 words.%
}{%
In the oxidative pentose phosphate pathway, one molecule of glucose-6-phosphate is converted to ribulose-5-phosphate, producing two molecules of NADPH and one molecule of CO2. ATP is not produced directly in this pathway. The two NADPH-producing enzymes are glucose-6-phosphate dehydrogenase (G6PD) and 6-phosphogluconate dehydrogenase (6PGD). This pathway supports proliferating cancer cells by providing NADPH, which is essential for maintaining cellular redox balance and supporting biosynthetic reactions necessary for cell growth and proliferation.%
}

\paragraph{Assessment and significance.}
The answer supplies all requested quantitative facts: two NADPH and one
CO$_2$ per glucose-6-phosphate, no direct ATP production, and both
NADPH-producing enzymes (G6PD and 6PGD). It connects these facts to redox
maintenance and biosynthesis, illustrating the ability to turn pathway
knowledge into a concise explanation of its relevance to proliferating cells.

\subsubsection{Example 2: Connecting a cancer mutation to epigenetic mechanisms}
\label{app:capability_example_2}

\qapair{%
What new enzymatic activity does the cancer-associated IDH1 R132H mutant acquire, which oncometabolite does it produce, and how can that metabolite alter DNA and histone methylation and cellular differentiation? Keep the answer under 150 words.%
}{%
The cancer-associated IDH1 R132H mutant acquires a new enzymatic activity that converts 2-oxoglutarate to the oncometabolite 2-hydroxyglutarate (2-HG). This metabolite can inhibit the activity of 2-oxoglutarate-dependent dioxygenases, including histone demethylases and DNA demethylases. As a result, 2-HG can alter DNA and histone methylation patterns, leading to changes in gene expression and cellular differentiation. Specifically, 2-HG can inhibit the activity of histone demethylases, such as KDM5C, which are involved in the removal of methyl groups from histone lysine residues. This inhibition can lead to the accumulation of methylated histones and altered chromatin structure, affecting gene expression and cellular differentiation. Additionally, 2-HG can also inhibit the activity of DNA demethylases, such as TET1 and TET2, which are involved in the removal of methyl groups from DNA. This inhibition can lead to the accumulation of methylated DNA and altered gene expression, further contributing to the oncogenic phenotype.%
}

\paragraph{Assessment and significance.}
The answer connects the mutant enzyme's conversion of 2-oxoglutarate to
2-HG with inhibition of dioxygenases, increased methylation, and altered
gene expression and differentiation. This mechanistic chain is useful for
explaining how a metabolic alteration can affect cell state. The core
mechanism is correct. Its description of TET1/TET2 as DNA demethylases is
functional shorthand for their role as 5-methylcytosine oxidases in DNA
demethylation.

\subsubsection{Example 3: Interpreting luminal epithelial and mesenchymal programs}
\label{app:capability_example_3}

\qapair{%
Compare the baseline expression profiles of MCF7 and MDAMB231 using ESR1, GATA3, FOXA1, CDH1 and VIM. For each gene, state which profile has higher expression. Then explain what the combined pattern suggests about luminal epithelial versus mesenchymal programs. Describe relative expression rather than claiming absolute absence. Keep the answer under 150 words.%
}{%
ESR1, GATA3, and FOXA1 show higher expression in MCF7, while CDH1 is elevated in MCF7 and VIM is elevated in MDAMB231. This pattern indicates that MCF7 maintains a luminal epithelial program characterized by hormone receptor signaling and epithelial adhesion, whereas MDAMB231 exhibits a mesenchymal program marked by intermediate filament expression. The contrast between these profiles reflects distinct cellular identities, with MCF7 retaining differentiated epithelial features and MDAMB231 displaying markers associated with mesenchymal transition.%
}

\paragraph{Assessment and significance.}
All five expression directions match the recorded DepMap values:
ESR1, GATA3, FOXA1, and CDH1 are higher in MCF7, whereas VIM is higher in
MDAMB231. The answer integrates the markers into a coherent interpretation
of luminal epithelial versus mesenchymal programs and describes relative
expression without claiming absolute absence. This is useful for translating
gene-level comparisons into interpretable cell-state summaries; the baseline
comparison supports a mesenchymal-associated state, not an observed
transition over time.

\subsubsection{Example 4: Structured comparison of myeloid and epithelial lineages}
\label{app:capability_example_4}

\qapair{%
Compare the baseline profiles of THP1 and A549. Which has higher expression of each of these genes: LYZ, LST1, TYROBP, FCER1G, SPI1, KRT8 and KRT18? Give a table with gene and higher-expression cell line, followed by one sentence interpreting the lineage difference. Do not discuss treatment effects.%
}{%
\begin{center}
\begin{tabular}{ll}
\toprule
gene & higher-expression cell line \\
\midrule
LYZ & THP1 \\
LST1 & THP1 \\
TYROBP & THP1 \\
FCER1G & THP1 \\
SPI1 & THP1 \\
KRT8 & A549 \\
KRT18 & A549 \\
\bottomrule
\end{tabular}
\end{center}
The data delineates a clear myeloid versus epithelial lineage boundary, with THP1 expressing canonical monocyte markers and A549 expressing simple epithelial keratins.%
}

\paragraph{Assessment and significance.}
All seven expression directions match the recorded DepMap values, and
the response follows the requested table-plus-one-sentence format without
introducing treatment effects. It combines five myeloid-associated markers
with two epithelial keratins to explain the lineage contrast. This example
illustrates a compact, auditable gene-level comparison paired with a
biological interpretation, a useful format for reviewing sample identity
and communicating expression differences.

Examples 5--9 were run with the same checkpoint and greedy decoding, with the
reasoning mode enabled; each response reproduces the generated reasoning followed by
the concluding line. Cell-line names were resolved to their DepMap profiles as above.
Sample identifiers of CancerModels.org specimens and GEO samples were replaced by the
corresponding omics profiles and removed from the text, so the model received no
textual cue about the sample beyond the wording of the question. Training exposure
differs between the examples: the cell lines in Example 5 are excluded from all
gene-essentiality training data, although their profiles appear in other cell-line
tasks; the CancerModels.org specimen in Example 6 is part of the training data for
that source; and the studies in Examples 7--9 come from the publication-disjoint test
split of GEO-OmicsQA. All expression values quoted in the assessments are TPM from the
source matrices.

\subsubsection{Example 5: Inferring an oncogene dependency from receptor overexpression}
\label{app:capability_example_5}

\qapair{%
We are considering HER2-directed therapy for oesophageal squamous carcinoma. Given the profiles of TE4 and KYSE150, is ERBB2 essential for the survival of each line? Point to the feature in the profiles that best explains the difference, and say what it implies about the underlying genomic event.%
}{%
The expression data reveals a stark contrast in ERBB2 abundance between the two lines. KYSE150 exhibits only baseline transcript levels, whereas TE4 displays a massive upregulation of the receptor. This profound overexpression in TE4 implies that the cell has acquired a genomic alteration, such as an amplification, driving a heavy reliance on HER2 signaling. Consequently, ERBB2 is likely essential for TE4 survival, while it remains non-essential in KYSE150.

\textit{Concluding line:} Yes%
}

\paragraph{Assessment and significance.}
Both dependency calls match DepMap 24Q4: ERBB2 is essential in TE4 (dependency
probability 1.00, Chronos $-2.02$) and not in KYSE150 (0.06, Chronos $-0.16$), and
ERBB2 expression differs about 67-fold (2{,}045 versus 30 TPM). The inferred
amplification is supported by the profile itself: neighbouring genes of the 17q12
amplicon are co-elevated only in TE4 (MIEN1 1{,}524, GRB7 495, PGAP3 463 TPM, versus
15--44 TPM in KYSE150), although the answer does not cite them. Both lines are held
out from the essentiality training split. The example links receptor
abundance to oncogene addiction and to a probable copy-number event within one
tissue type, the reasoning needed to prioritize HER2-directed therapy.

\subsubsection{Example 6: Distinguishing a patient tumour from xenograft-derived material}
\label{app:capability_example_6}

\qapair{%
Specimen 216738-123-R comes from the same renal cell carcinoma model. Classify it as a patient tumour, a xenograft, an organoid or a cell line, and justify the call in three bullet points naming immunological cell markers, antigens and other relevant markers%
}{%
\begin{itemize}
\item SPP1, HLA-DRA, and C1QB are detected, providing a combined signal of myeloid-associated antigens and complement components that is characteristic of patient-derived tumour material.
\item The presence of HLA-DRB1 and C1QA further supports an immune-active microenvironment typical of human patient specimens, rather than the limited transcriptomic scope of cell lines or organoids.
\item This specific constellation of markers aligns with the profile of a patient tumour, distinguishing it from the more restricted expression patterns expected in xenografts, organoids, or cell lines.
\end{itemize}

\textit{Concluding line:} patient tumour%
}

\paragraph{Assessment and significance.}
The classification is correct: the specimen is the patient tumour of a
CancerModels.org renal cell carcinoma model. Four of the
five cited transcripts separate it from a xenograft of the same model
(216738-123-R-FEV): HLA-DRA (6{,}271 versus 111 TPM), HLA-DRB1 (4{,}555 versus 66),
C1QA (613 versus 0.3) and C1QB (505 versus 0.2). These genes mark human
antigen-presenting and complement-producing myeloid cells, which are present in
patient tissue but absent from xenografts, whose stroma is of mouse origin. SPP1 is
not discriminative, as it is equally high in the xenograft (8{,}231 versus 9{,}541 TPM)
and is therefore expressed by the tumour cells. The prompt names the marker classes to
report, which partly guides the answer. The example shows that the model can reason
about the tumour microenvironment as well as the malignant compartment, which is
relevant to quality control of patient-derived models.

\subsubsection{Real Study Examples}
\label{app:capability_real_studies}

Examples 7--9 are adapted from the held-out, publication-disjoint test split of
GEO-OmicsQA. Each prompt keeps the original question and appends instructions to
describe the observations and reasoning. The GSM identifiers were replaced by the
corresponding profiles, so the sample groups are defined only by their position and by
the wording of the question.

\subsubsection{Example 7: Host response to LCMV polymerase activity (GSE173072)}
\label{app:capability_example_7}

\qapair{%
We found that the LCMV L protein's ability to activate STAT3 is tied to its RNA-dependent RNA polymerase (RdRp) activity. Looking at the expression of DDIT3 and FOS in the RNA-seq data from samples GSM5259349, GSM5259350, GSM5259351 and GSM5259352, GSM5259353, GSM5259354, what does this suggest about the functional consequences of this RdRp activity on the host genome? State which sample group shows higher expression. Then check whether unfolded protein response genes (HSPA5, XBP1, ATF4, TRIB3, PPP1R15A), immediate-early genes (JUN, EGR1, ATF3), interferon-stimulated genes (ISG15, IFIT1) and STAT3 itself change in the same direction, and explain what kind of host response this pattern indicates. Answer in full sentences.%
}{%
The expression data reveals that DDIT3 and FOS are significantly upregulated in the second group of samples compared to the first. This induction is accompanied by a concurrent upregulation of unfolded protein response genes (HSPA5, XBP1, ATF4, TRIB3, PPP1R15A), immediate-early genes (JUN, EGR1, ATF3), and interferon-stimulated genes (ISG15, IFIT1). The coordinated activation of these diverse stress and immune pathways indicates that the RdRp activity of the viral protein triggers a massive, multi-faceted host stress response, likely driven by the aberrant activation of STAT3.

\textit{Concluding line:} The second group shows higher expression.%
}

\paragraph{Assessment and significance.}
The prompt does not state which group is infected, so the direction has to be read
from the profiles. The second group is the LCMV-infected one, consistent with its
interferon response (ISG15 about 690 versus 6 TPM, IFIT1 113 versus 0.3). DDIT3 rises
from 8 to 237 TPM and FOS from about 1 to 8. The unfolded protein response is induced
(HSPA5 4.3-fold, XBP1 3-fold, TRIB3 5-fold, PPP1R15A 13-fold), as are the
immediate-early genes (JUN 10-fold, EGR1 from 0 to 42 TPM, ATF3 from 2 to 291 TPM).
ATF4 changes only 1.2-fold, so its inclusion overstates the effect, and the answer
does not report STAT3 itself (1.8-fold); attributing the response to STAT3 follows the
premise of the question. The integrated reading, an endoplasmic-reticulum stress and
interferon response to viral RNA synthesis, matches the data and shows how the model
summarizes coordinated changes across several pathways.

\subsubsection{Example 8: Resisting a leading question on KDM2B knockdown (GSE81043)}
\label{app:capability_example_8}

\qapair{%
Evaluating the control glioblastoma samples (GSM2141218, GSM2141219) against the KDM2B knockdowns (GSM2141220, GSM2141221), are anti-apoptotic genes like MCL1 globally upregulated by the perturbation? Before answering, describe what you observe for the genes in question in each sample group, then explain the underlying biology step by step. Explain your reasoning, then answer with Yes or No on the last line.%
}{%
The expression data reveals that anti-apoptotic genes, including MCL1, are actually downregulated in the knockdown samples compared to the controls. This reduction in survival factors indicates that the perturbation compromises cellular viability rather than promoting it.

\textit{Concluding line:} No%
}

\paragraph{Assessment and significance.}
The answer rejects the premise of the question, correctly. The knockdown is effective
(KDM2B 13 versus 2 TPM), MCL1 falls from about 180 to 88 TPM and BCL2A1 from 9 to 1.
The generalization to anti-apoptotic genes as a class is partly overstated, since
BCL2L1, BIRC5 and XIAP are unchanged. The interpretation of reduced viability is
supported by pro-apoptotic genes that the answer does not mention: PMAIP1 (NOXA,
20 to 47 TPM), TNFRSF10B (DR5, 112 to 165) and CASP8 (18 to 29). The step-by-step
description requested in the prompt is brief. The example shows that the model follows
the data rather than the framing of a leading question.

\subsubsection{Example 9: DNA-repair gene loss after CDK12/13 degradation (GSE262119)}
\label{app:capability_example_9}

\qapair{%
Based on the RNA-seq data for DMSO control samples (GSM8157881, GSM8157882, GSM8157883) and YJ9069-treated samples (GSM8157884, GSM8157885, GSM8157886), is there a noticeable decrease in the expression of the DNA double-strand break repair genes BRCA1 and Rad51 upon treatment? Before answering, describe what you observe for the genes in question in each sample group, then explain the underlying biology step by step. Explain your reasoning, then answer with Yes or No on the last line.%
}{%
The expression data reveals a marked downregulation of BRCA1 and Rad51 in the YJ9069-treated samples compared to the DMSO controls. These genes are critical for homologous recombination, the primary pathway for repairing DNA double-strand breaks. The reduction in their transcript levels indicates that the treatment impairs the cell's ability to mount an effective DNA damage response, thereby compromising genomic stability.

\textit{Concluding line:} Yes%
}

\paragraph{Assessment and significance.}
Both genes decrease after treatment: BRCA1 from about 16 to 4 TPM and RAD51 from 11
to 2. Other long DNA-damage-response genes follow the same pattern (ATR 6 to 0, FANCD2
29 to 7, FANCI 61 to 21 TPM), in line with the known role of CDK12/13 in transcriptional
elongation of long genes and the resulting homologous-recombination deficiency; the
answer states the functional consequence but not this mechanism. Housekeeping genes
such as GAPDH and ACTB roughly double in these TPM profiles, reflecting a compositional
shift, but the four- to six-fold decrease of BRCA1 and RAD51 exceeds it. The example
connects a drug-induced expression change to a DNA-repair phenotype relevant to
therapeutic combinations with PARP inhibition.

\subsection{Multi-turn Conversation about Epithelial and Mesenchymal Cell States}
\label{app:multiturn_conversation}

This five-turn conversation illustrates how \modelname{} connects an initial
expression comparison to follow-up questions about cell identity, causal
inference, and a hypothetical perturbation experiment. We used the model with greedy decoding,
no system message, and a maximum of 16,000 new tokens per turn.
The chat interface inserted the two corresponding 20,541-feature omics
profiles into the first prompt; conversation history and those profile
inputs were retained throughout, without resetting between turns.
The user-entered prompts and model responses below are reproduced verbatim,
with only LaTeX formatting applied. The observations in the final prompt
are hypothetical, not results of a laboratory experiment.

\subsubsection{Turn 1: Comparing baseline expression profiles}

\qapair{%
Compare the baseline expression profiles of MCF7 and MDAMB231 using ESR1, GATA3, FOXA1, CDH1 and VIM. Which profile has higher expression of each gene, and what biological programs does the pattern suggest? Keep your answer under 130 words.%
}{%
ESR1, GATA3, FOXA1, and CDH1 are highly expressed in MCF7, while VIM is elevated in MDAMB231. This pattern indicates that MCF7 is a luminal epithelial breast cancer cell line, whereas MDAMB231 is a mesenchymal-like cell line.%
}

\subsubsection{Turn 2: Distinguishing cell state from a historical transition}

\qapair{%
Does that comparison demonstrate that one of these cell lines underwent epithelial-to-mesenchymal transition, or that it arose from the other? Explain what we can and cannot infer from these baseline profiles. Keep your answer under 100 words.%
}{%
No, the comparison does not demonstrate that one cell line underwent epithelial-to-mesenchymal transition or arose from the other. The baseline profiles only reflect the current state of the two cell lines, not their evolutionary history.%
}

\subsubsection{Turn 3: Connecting adhesion to cell identity}

\qapair{%
Focusing on the adhesion gene in our comparison: what does its protein do, and would restoring that gene alone in the more mesenchymal line necessarily recreate the luminal identity of the other line? Explain briefly, in under 130 words.%
}{%
CDH1 encodes E-cadherin, a calcium-dependent cell adhesion protein that mediates homophilic interactions to maintain epithelial integrity. Restoring CDH1 alone in the mesenchymal line would not necessarily recreate the luminal identity, as other luminal-specific genes like ESR1, GATA3, and FOXA1 are also required.%
}

\subsubsection{Turn 4: Outlining a perturbation experiment}

\qapair{%
Suppose we want to test whether TGF-beta can induce an EMT-like change in the more epithelial line. Propose a small time-course experiment with controls, molecular readouts and one functional readout. Explain how you would avoid mistaking cell growth for migration. Treat the response to TGF-beta as something to test, not assume. Keep your answer under 180 words.%
}{%
Treat MCF7 cells with TGF-beta and monitor changes in VIM expression and cell migration over time. Use untreated MCF7 cells as a control. Measure cell growth separately to ensure that any observed changes in migration are not due to differences in cell growth.%
}

\subsubsection{Turn 5: Interpreting mixed hypothetical results}

\qapair{%
Here is a hypothetical result from that experiment: VIM RNA increases, CDH1 RNA is unchanged, junctional E-cadherin staining becomes weaker, and a scratch closes faster, but treated cells also proliferate faster. How would you interpret this mixed result, and which two follow-up measurements would most help resolve it? Keep your answer under 150 words.%
}{%
The mixed result suggests that TGF-beta treatment induces partial EMT-like changes, including increased VIM expression and weakened junctional E-cadherin staining, but does not completely abolish CDH1 expression. The faster scratch closure could be due to both increased cell migration and faster cell proliferation. To resolve this, I would measure cell proliferation rates and assess the functional impact of TGF-beta treatment on cell migration in the absence of proliferation.%
}

\paragraph{Assessment and significance.}
All five expression directions in the first response agree with the cached
DepMap source: ESR1, GATA3, FOXA1, and CDH1 are higher in MCF7, whereas VIM
is higher in MDAMB231. The follow-ups retain the relevant biological context,
distinguish a baseline phenotype from evidence of a historical transition,
and connect CDH1 to E-cadherin-mediated adhesion without equating restoration
of one gene with recreation of a luminal identity. The final response
recognizes that proliferation can contribute to faster scratch closure.

\end{document}